\documentclass[twocolumn,tighten,times]{aastex6}

\turnoffedit

\shorttitle{Schad et al.}
\shortauthors{Schad et al.}

\usepackage{enumitem}
\newcommand{\helix}{\textsc{HeLIx$^\textbf{+}$}}
\newcommand{\heI}{He \textsc{i}{\ }}

\begin{document}

\title{Vector magnetic field measurements along a cooled stereo-imaged coronal loop}

\author{T.A. Schad\altaffilmark{1}, M.J. Penn\altaffilmark{2}, H. Lin\altaffilmark{3}, and P.G. Judge\altaffilmark{4}}
\affil{	\altaffilmark{1} National Solar Observatory$^{*}$, 8 Kiopa`a Street, Pukalani, HI 96768, USA; \url{schad@nso.edu}  \\
	\altaffilmark{2} National Solar Observatory$^{*}$, 950 N. Cherry Ave., Tucson, AZ 85719, USA \\
       	\altaffilmark{3} Institute for Astronomy, University of Hawai`i, Pukalani, HI 96768, USA \\
       	\altaffilmark{4} High Altitude Observatory, National Center for Atmospheric Research$^{**}$, P.O. Box 3000, Boulder, CO 80307-3000, USA}

\altaffiliation{$^{*}$ The National Solar Observatory (NSO) is operated by the Association of Universities for Research in Astronomy, Inc. (AURA), under cooperative agreement with the National Science Foundation.}	
\altaffiliation{$^{**}$ The National Center for Atmospheric Research is sponsored by the National Science Foundation.}

\begin{abstract}
The variation of the vector magnetic field along structures in the solar corona remains unmeasured.  Using a unique combination of spectropolarimetry and stereoscopy, we infer and compare the vector magnetic field structure and three-dimensional morphology of an individuated coronal loop structure undergoing a thermal instability.  We analyze spectropolarimetric data of the \heI 10830 \mbox{\AA} triplet ($1s2s{\ }^{3}S_{1} - 1s2p{\ }^{3}P_{2,1,0}$) obtained at the Dunn Solar Telescope with the Facility Infrared Spectropolarimeter on 19 September 2011.  Cool coronal loops are identified by their prominent drainage signatures in the \heI data (redshifts up to 185 km sec$^{-1}$).  Extinction of EUV background radiation along these loops is observed by both the Atmospheric Imaging Assembly onboard the \textit{Solar Dynamics Observatory} and the Extreme Ultraviolet Imager onboard spacecraft A of the \textit{Solar Terrestrial Relations Observatory}, and is used to stereoscopically triangulate the loop geometry up to heights of 70 Mm ($0.1$ $R_{sun}$) above the solar surface.  The \heI polarized spectra along this loop exhibit signatures indicative of atomic-level polarization as well as magnetic signatures through the Hanle and Zeeman effects.  Spectropolarimetric inversions indicate that the magnetic field is generally oriented along the coronal loop axis, and provide the height dependence of the magnetic field intensity.  The technique we demonstrate is a powerful one that may help better understand the thermodynamics of coronal fine structure magnetism.
\end{abstract}



\section{Introduction}

The solar corona is a low beta\footnote{The plasma beta parameter is given by the ratio of the kinetic pressure to the magnetic pressure.}, highly conducting magnetoplasma whose evolution is controlled by the magnetic field coupling it to the Sun and its connections to interplanetary space.  Various techniques to remotely measure the magnetic field of the hot corona are available in principle (as reviewed by \citet{judge2001}) but the observational challenges are severe.  Here, we take advantage of coronal material that has cooled to temperatures where significant populations of neutral He atoms exist, and apply spectropolarimetry of the \heI triplet.  Having sensitivity to both the Zeeman and Hanle effects \citep{trujillo_bueno_2002,asensio_ramos2008}, the \heI triplet is a powerful magnetic diagnostic of solar plasma in chromospheric thermal conditions (\textit{e.g.}, prominences \citep{casini2003,casini2009}, spicules \citep{centeno2010}, filaments \citep{kuckein2012mag}, and active region superpenumbrae \citep{schad2013,schad2015}).  

Renewed attention is being given to cool material in the active corona.  High-resolution spatio-temporal imaging in Ca \textsc{ii} H \citep{antolin2010} and H${\alpha}$ \citep{antolin2012sharp} has revealed coronal rain---\textit{i.e.}, the cool, partially ionized, material that condenses from the hot corona into elongated blobs, clumps and streams, and subsequently falls towards the solar surface---to be more abundant than once thought, with observed mass drainage rates only a few factors less than the estimated mass flux into the corona by spicules.  The rain blobs are intricately structured, have widths typically 600 km or less, lengths of 2 Mm or less, temperatures of $10^4$ to $10^5$ K, and inhomogeneous densities \citep{antolin2015}.  Its formation is linked to thermal instabilities which may offer a discriminant for coronal heating processes \citep{field1965, antolin2010}. Numerical studies are now pushing to understand the observed rain dynamics using fully-ionized 3D MHD simulations \citep{moschou2015}, as well as partially-ionized hydrodynamic simulations \citep{oliver2016}.  Still, the macro-scaled and fine-scaled magnetic structure of these draining loops remains largely unconstrained by observations.

Using spectropolarimetry of the \heI 10830 \mbox{\AA} triplet, this work studies the magnetic field structure of a transient, rapidly down-flowing, loop-like coronal feature---henceforth referred to as a `coronal loop'---rooted within a sunspot, and identified as a variant of the coronal rain phenomena.  Our observations provide a unique combination of on-disk \heI spectropolarimetry and stereoscopic EUV imaging, which allow us to directly compare the stereoscopically triangulated loop topology with the measured \heI polarization signals and inverted magnetic field parameters, up to heights of 70 Mm (\textit{i.e.}, $\sim 0.1 R_{\Sun}$) above a solar active region.  


\begin{figure*}
\centering
\includegraphics[width=0.9\textwidth]{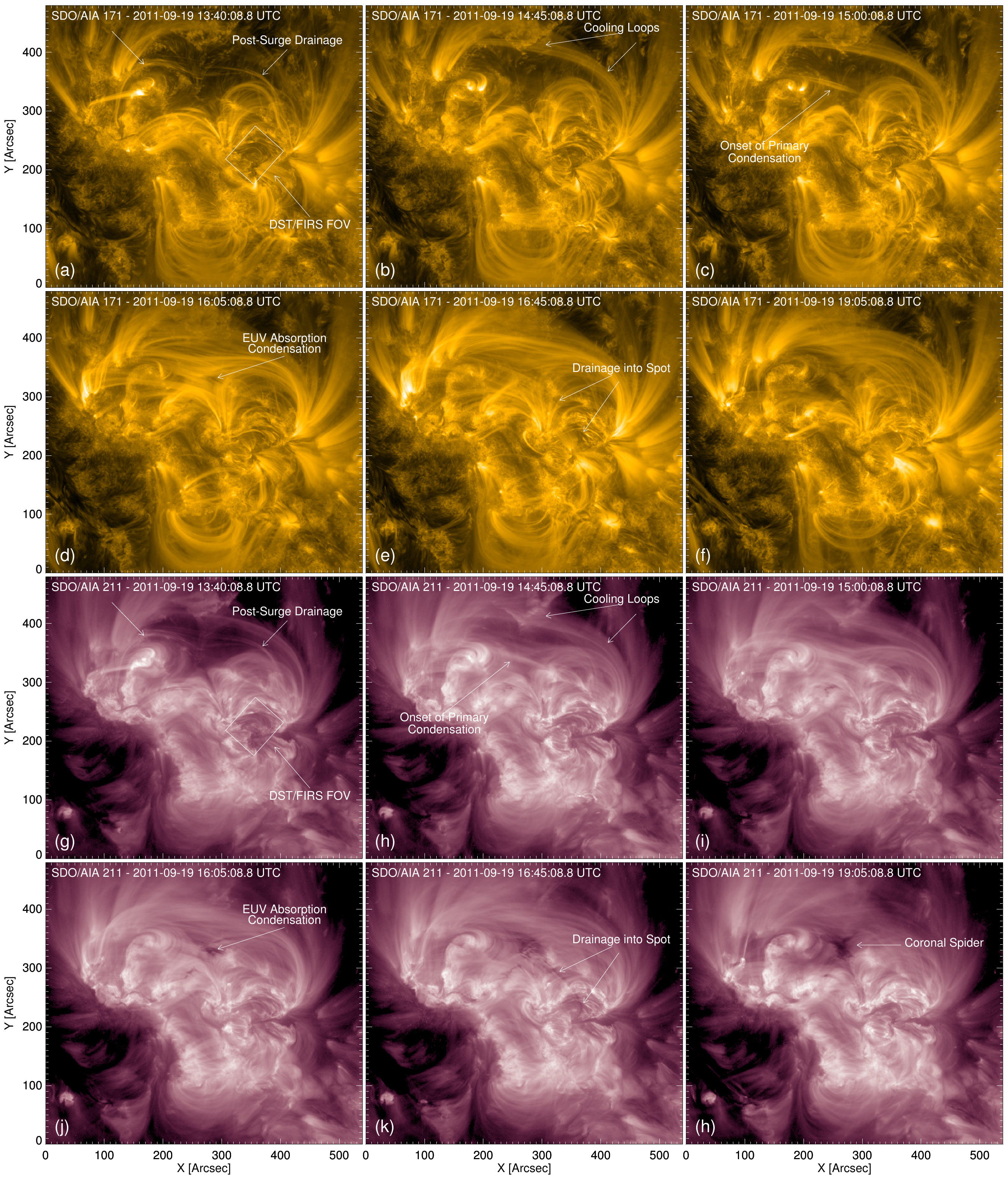}
\caption{Temporal evolution of the active region observed on 19 September 2011, as viewed by the SDO/AIA 171 \mbox{\AA} (panels a - f) and 211 \mbox{\AA} (panels g-h) channels.  The selected time steps demonstrate the cooling and condensation of the overlying loop system as discussed in the text.  Spatial coordinates are held fixed relative to the region, for which solar rotation has been compensated.  Solar North is up in the figure, while the DST/FIRS scanned field-of-view (\textit{solid outlined box}) is identified in panels (a) and (g).  The dark drainage channel identified in panels (e) and (k) corresponds to the coronal loop studied in this work. \\ 
\\
(An animation and a color version of this figure are available online)}. 
\label{fig:aia_images}
\end{figure*}

\section{Observations}\label{sec:data_descr}

The observed coronal loop, which we select and analyze in detail in Sections~\ref{sec:methods} and~\ref{sec:pol_analysis}, is one of multiple coronal drainage features observed during a large-scale cooling event between 14:45 and 17:50 UTC within the multi-polar complex above NOAA active regions 11296, 11295, and 11298 on 19 September 2011.


\subsection{EUV Temporal Evolution}

Figure~\ref{fig:aia_images} presents a contextual overview of the observed region, the cooling event, and subsequent drainage features, using EUV data from the Atmospheric Imaging Assembly \citep[SDO/AIA:][]{lemen2011} instrument onboard NASA's Solar Dynamics Observatory \citep[SDO:][]{pesnell2012}.  We downloaded and examined the 12 second cadence, level 1 AIA data, in each of the 304, 131, 171, 193, 211, and 335 \mbox{\AA} EUV channels between 12:00 and 20:00 UTC.  Each level 1, full-disk, data file is processed using \textit{aia\_prep} v5.1, establishing a common coordinate system and $0.6''$ per pixel spatial sampling.  We manually verified the alignment between spectral channels before cutting out a $540'' \times 480''$ subregion in each file surrounding the targeted region centered at N23W11 at 17:04 UTC ($\cos \theta = \mu = 0.92$), which we track according to the mean synodic solar rotation rate \citep{howard1990}.

Between 14:45 and 17:50 UTC, the active region complex shown in Figure~\ref{fig:aia_images} exhibits a progressive illumination of cooler EUV channels (seen in all AIA channels) indicative of large scale ``catastrophic cooling'' in the corona \citep{kjeldseth_moe_1998, schrijver2001, kamio2011}.  A time lag of approximately 15 minutes, for example, is evident between the onset of the ``primary condensation" in 211 \mbox{\AA} (14:45:08.8 UTC; panel (h)) relative to 171 \mbox{\AA} (15:00:08.8 UTC; panel (c)), which have characteristic temperatures of 2 MK and 0.63 MK, respectively \citep{lemen2011}.  By 15:00 UTC, many more coronal loops are apparent in 211 \mbox{\AA} than in 171 \mbox{\AA}, as the cooling continues.  

Catastrophic cooling often triggers condensation within the corona.  When viewed off-limb, the resulting cool material usually appears as intensity brightenings in 304 \mbox{\AA} observations, as well as in spectral lines like Ca II K, Ca II 8542 \mbox{\AA}, and H$\alpha$.  Condensations are much harder to observe on-disk; though, coronal rain has been observed previously using on-disk visible-light spectroscopy \citep{antolin2012ondisk,ahn2014}.  In the EUV channels, cool, condensed material on-disk can be identified by its absorption of background EUV emission due to the photoionization of its neutral H$^{0}$, neutral He$^{0}$, and singly-ionized He$^{+1}$ material.  In Figure~\ref{fig:aia_images}, we find cool material beginning to condensate just prior to 16:05 UTC (see, e.g., panel (j)) near the ``primary condensation'' site identified in panels (c) and (h).  This site is associated with the loop apex due to its mid-point location between the system outer footpoints.  The enhanced contrast of the cool material in the 211 \mbox{\AA} images relative to the 171 \mbox{\AA} is consistent with the wavelength dependence of the photoionization cross sections, which general increase towards longer wavelengths, up until the He$^{+}$ ionization cutoff bound at 911 \mbox{\AA} \citep{gilbert2011}.

Shortly after the cool material begins to form at the system apex, blobs and loop-like streams of cool material begin to show lateral movement away from the cool concentration at the apex along curvilinear loops towards a large sunspot centered at $(X,Y)= (360'',230'')$.  This suggests that some portion of cool material at the apex becomes gravitational unstable and drains towards the solar surface, which by definition is coronal rain. This is confirmed by the \heI observations presented in the following sections, wherein the dark (cool) coronal loop feature of panel (k) is analyzed further. 

Despite large scale drainage occurring between 16:45 and 17:20 UTC, the cool material continues to accumulate and becomes darker near the apex of the observed region.  By 19:05 UTC, a persistent cool feature is present at the system apex that is consistent with a coronal cloud filament (also known as a coronal spider) due to the high formation height of the structure identified below with stereoscopy \citep{allen1998, lin_martin2006}.  The filament lasts another 7 hours, during which a number of ``spider legs'' (i.e. curvilinear drainage channels extending from the filament) form.  


\begin{figure*}
\centering
\includegraphics[width=0.85\textwidth]{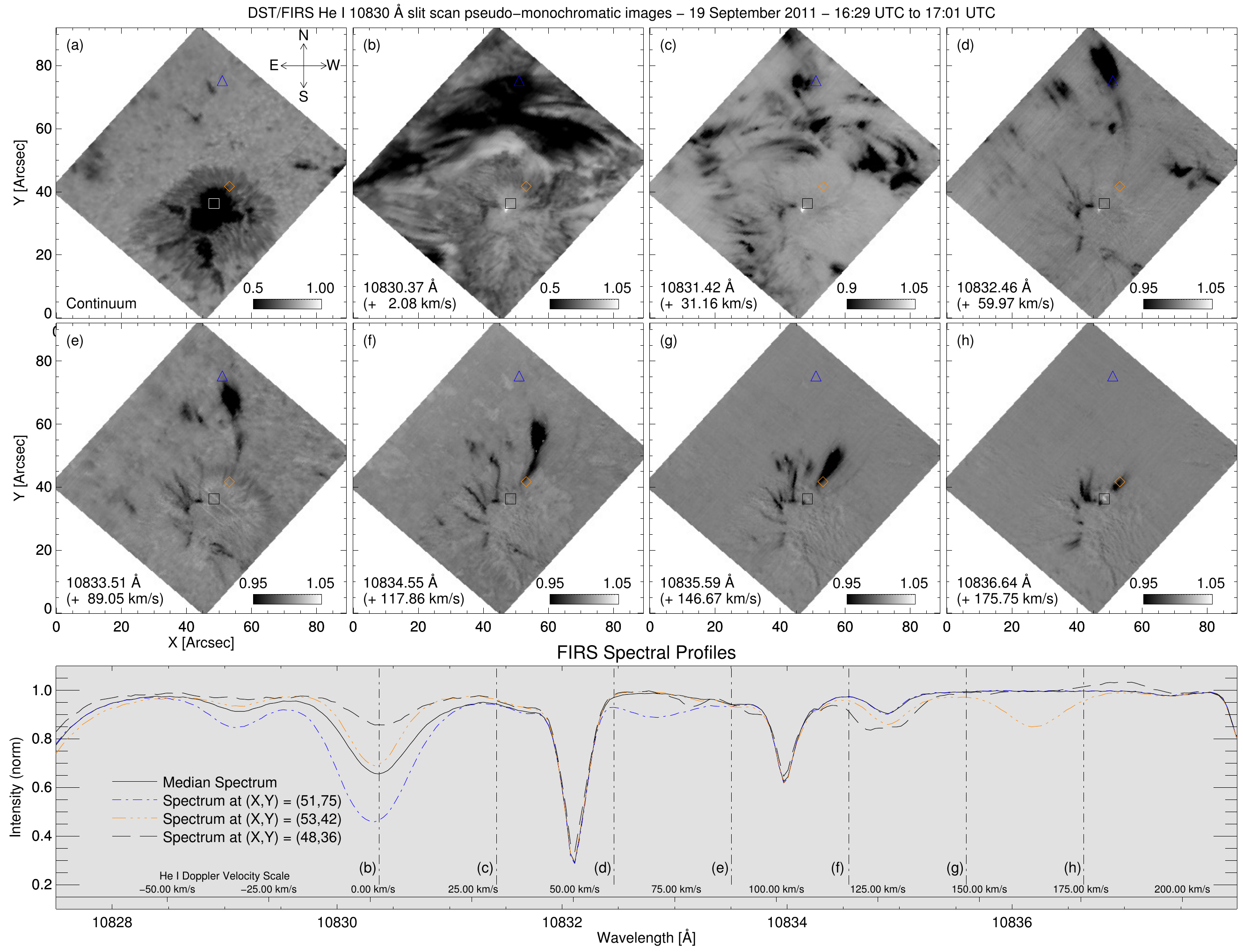}
\caption{Overview of DST/FIRS coronal rain observations on 19 September 2011.  \textit{(a)} Continuum intensity map of slit-scanned field-of-view containing the large sunspot of AR 11298. \textit{(b-h)} Pseudo-monochromatic slices of the data cube at wavelengths ranging from 10830.37 \mbox{\AA} to 10836.64 \mbox{\AA}, \textit{i.e.,} Doppler redshifts of 2.08 to 175.75 km/sec relative to the average wavelength of the \heI red transitions (10830.295 \mbox{\AA}, see Section~\ref{sec:heI_pol_sig}).  (\textit{Bottom}) Spectral line profiles extracted from the locations indicated in panels a-h, illustrating the presence of additional \heI velocity components redward of the telluric feature at 10832.1 \mbox{\AA}, and relative to the median spectrum calculated using the entire data cube.  A \heI Doppler velocity scale bar is added for reference. \\ 
\\
(A color version of this figure are available online)}
\label{fig:firs_obs}
\end{figure*}

\subsection{\heI Spectropolarimetry}\label{sec:obs_firs}

The signature of the cool, draining, coronal material in our neutral helium observations is the presence of an additional velocity component of the \heI 10830 \mbox{\AA} triplet that exhibits acceleration along loop-like paths (see Figure~\ref{fig:firs_obs}).  These \heI spectropolarimetric data were obtained during the catastrophic cooling event described above using the Facility Infrared Spectropolarimeter \citep[FIRS:][]{jaeggli2010} at the NSO's 0.76 m Dunn Solar Telescope. 

FIRS is a long-slit scanning, diffraction-grating based, full-Stokes, dual-beam spectropolarimeter.  Here we examine a single FIRS scan of the large sunspot within NOAA AR 11298, with a field-of-view of $60'' \times 70''$, as indicated in Figures~\ref{fig:aia_images} and~\ref{fig:firs_obs}.  The data have a spatial sampling of $0.29''$ and an observed spectral range of 35.743 \mbox{\AA} ($\Delta\lambda$ = 38.683 m\mbox{\AA} pix$^{-1}$).  The slit was oriented at a $41.3^{\circ}$ angle with respect to the solar meridian.  Beginning at 16:29 UTC, the slit was scanned from solar northeast to southwest with a step size equal to the projected slit width of $0.29''$. At each slit position, spectra were acquired using a sequence of four efficiency-balanced modulation states using 125 msec exposures, which was repeated once for a total of 1 second cumulative integration time.  The 218 step scan lasted 32 minutes, ending at 17:01 UTC, thus bracketing in time the EUV drainage described above (\textit{i.e.}, Figure~\ref{fig:aia_images} panels (e) and (k)).

Standard reduction methods were applied to the FIRS data in the manner described by \cite{schad2013}, which includes dark and flat calibrations, geometric registration, wavelength calibration, and polarimetric crosstalk removal.  One exception here is that the spectral dispersion was determined by fitting the line cores of the two telluric lines at $10833.981 \mbox{\AA}$ and  $10832.108 \mbox{\AA}$, as in \cite{kuckein2012}.  Residual interference fringes were removed from the polarized spectra using the 2D pattern recognition technique of \cite{casini2012_fringes}, resulting in an un-binned mean noise level in Stokes (Q,U,V) of $(1.0,0.92,1.3) \times 10^{-3}$, in units of the incident intensity.  The reference direction for Stokes +Q is parallel to the solar equator.  

Despite real-time seeing correction and image stabilization by the DST high-order adaptive optics system \citep{rimmele2004}, the solar image experienced shifts up to $0.1''$ between adjacent FIRS slit positions due to short periods of degraded atmospheric seeing, with a cumulative temporal drift of $2.55''$ during the scan.  These shifts have been corrected through careful image registration with a co-operating G-band context imager.  Image shifts determined from the G-band images are converted to their equivalent shift within the FIRS map.  Any spatial maps of feature topology or physical quantities derived from the FIRS maps are corrected using Delaunay triangulation of the translated observation points \citep{lee1980}, and subsequent quintic-polynomial smooth interpolation.  

Figure~\ref{fig:firs_obs} panels b-h show slices of the observed data cube at constant, and progressively longer (\textit{i.e.}, redshifted relative to the rest wavelengths of \heI), wavelengths (panels b through h), which sample Doppler shifted neutral helium material progressively between 0 and 175 km/sec.  The data in the figure have been normalized to the local continuum intensity.  At shorter (less redshifted) wavelengths, the chromosphere appears densely structured with low-lying loops and fibrils, which are oriented primarily in the east-west direction within the region north of the sunspot umbra.  At longer (more redshifted; $>50 km/s$) wavelengths, these low-lying structures are no longer visible, but instead dark, curvilinear structures with a more north-south orientation begin to appear, first in the northern portion of the image and then progressively closer to the sunspot umbra.  These loop-like structures are the \heI triplet counterpart to the EUV absorption discussed above, and give evidence that cool material is draining from coronal heights along these loops, and furthermore that this material accelerates as it approaches the solar surface.  

The footpoints of the draining loops are located near the inner penumbral boundary of the sunspot.  At the footpoint of the most prominent (largest absorption) and highest redshifted coronal loop (marked with a square in Figure~\ref{fig:firs_obs}) velocities approach 185 km/sec, above the sound speed in the lower chromosphere.  The \heI profiles switch, at these speeds, from absorption to emission indicating significant local heating, likely due to a shock \citep[see, e.g.,][]{cargill1980}.  The component of \heI that is not highly redshifted also exhibits an emission profile at the loop footpoint (see panels b and e of Figure~\ref{fig:firs_obs}).  While this shock is also of interest, below we concentrate on the polarized signatures measured by FIRS along these coronal rain features. 


\subsection{Stereoscopic Observations}

Without stereoscopic observations, the above observations are blind to the vertical structure of the active region and the geometry of the coronal loops are poorly constrained.  Thankfully, on this date, spacecraft A of the twin spacecraft Solar Terrestrial Relations Observatory \citep[STEREO:][]{kaiser2005} was located in its eccentric orbit $103.67^{\circ}$ ahead of Earth, at a distance of $0.966$ AU from the Sun.  Meanwhile, SDO orbited Earth at $1.0046$ AU.  We examined coronal images taken in the 304 \mbox{\AA} and 195 \mbox{\AA} channels by the SECCHI/EUVI instrument onboard STEREO-A, acquired on 10 and 5 minute cadences, respectively.  The EUVI-A plate scale for this date and location was 1.59'' per pixel.  All STEREO/SECCHI/EUVI-A images were processed with the SolarSoft routines \textit{secchi\_prep} and \textit{euvi\_point}.  In Section~\ref{sec:stereo_recon} we use the 195 \mbox{\AA} observations, together with SDO/AIA 193 \mbox{\AA} observations, to triangulate the three dimensional geometry of the selected coronal structure. 


\section{Methods}\label{sec:methods}


\subsection{Data co-alignment}

The co-alignment of DST/FIRS and SDO/AIA data relies upon an anchor of co-temporal photospheric imaging observations acquired by both the DST and SDO.  We used the cross-correlation techniques implemented within the SSWIDL routine \textit{auto\_align\_images} to derive a general polynomial warping transformation between respective images used for alignment.  Facility alignment target observations were first used to co-register DST/FIRS data with the DST G-band imager, which covered a field-of-view of $87'' \times 87''$ with a plate scale of $0.0428''$ per pixel.  A single DST G-band image acquired during the FIRS scan and centered on the NOAA AR 11298 sunspot was then co-registered with a co-temporal SDO Helioseismic and Magnetic Imager (HMI) full-disk intensitygram, which was first processed up to level 1.5 using \textit{aia\_prep} v5.1.  These two derived transformations allow us to map the DST/FIRS observations into the corresponding SDO/HMI level 1.5 coordinate frame.  We find an additional scalar shift of $1.8''$ between the transformed DST/FIRS data and the SDO/HMI data is required to compensate for terrestrial atmospheric refraction.  We estimate that the accuracy of this alignment is better than $0.6''$ (\textit{i.e.}, the SDO level 1.5 pixel pitch); however, the overall alignment between DST/FIRS and the SDO/AIA data is limited by the alignment accuracy of the SDO level 1.5 data.  While \textit{aia\_prep} v5.1 uses the most up-to-date SDO master pointing files to establish a common geometry between SDO/HMI and SDO/AIA, spacecraft jitter and flexure still limit the pointing accuracy to $\lesssim 1.18''$ \citep{orange2014}. This is the primary source of alignment error in our analysis. 


\begin{figure}
\centering
\includegraphics[width=0.4\textwidth]{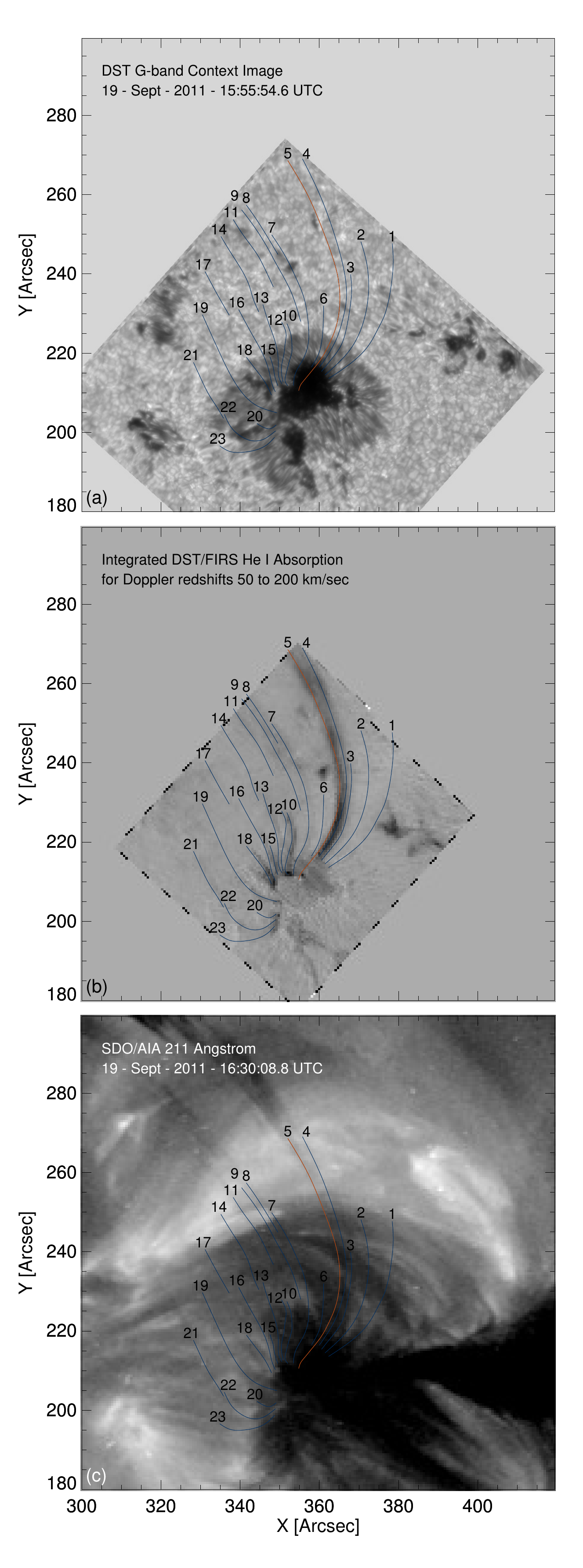}
\caption{Twenty-three coronal rain drainage loops, exhibiting redshifts greater than 50 km/sec, are manually traced by inspection of the entire DST/FIRS spectra data using {\sc Crispex} and coregistered with (a) the DST G-band context image and (c) the SDO/AIA data set here represented by the 211 \mbox{\AA} channel.  Panel (b) displays the integrated \heI absorption for all \heI spectral profiles of redshifts greater than 50 km/sec.  The loops of greatest absorption are \#4 and \#5. Loop \#5 (solid orange line) is analyzed in detail in Section~\ref{sec:pol_analysis}.  Spatial coordinates reference the axes shown in Figure~\ref{fig:aia_images}.\\ 
\\
(A color version of this figure and an animation of panel (c) are available online)}. 
\label{fig:firs_traces}
\end{figure}

\begin{figure*}
\centering
\includegraphics[width=0.85\textwidth]{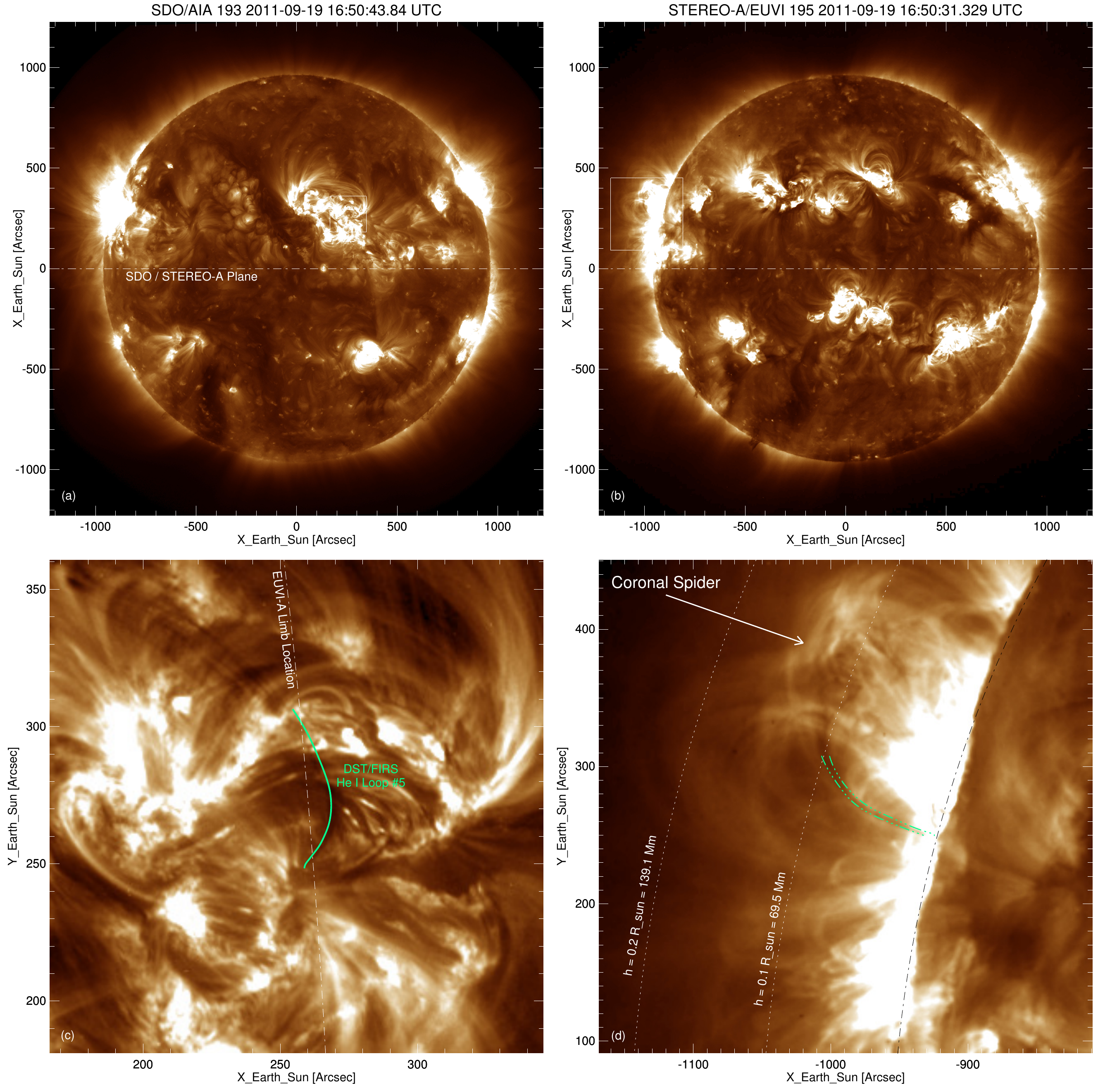}
\caption{Stereoscopic triangulation of the selected coronal loop. (\textit{Panels a-b}) SDO/AIA 193 and STEREO/EUVI-A 195 full-disk images rescaled and transformed into a common epipolar observational geometry.  The mid-plane of the image corresponds to the plane formed by the two spacecrafts.  (\textit{Panels c-d}) Magnifications of the subregions identified in the full-disk images.  The solid green line in panel (c) marks the loop \#5 position in the DST/FIRS data cube overplotted on the SDO/AIA image.  Green dot-dashed lines in panel (d) trace the corresponding \edit1{extinction, \textit{i.e.} darker, loop} feature off-limb in the STEREO/EUVI-A image. \edit1{This loop is the only prominent extinction feature extending to the surface, and is consistent with the large helium absorption observed by DST/FIRS along loop \#5.}  Spatial coordinates give angular offsets from disk center observed from the vantage of each spacecraft scaled to the Earth-Sun distance on this date. \\
\\
(A color version of this figure is available online)}. 
\label{fig:stereo_fig}
\end{figure*}

\subsection{DST/FIRS \heI Loop Tracing}\label{sec:trace_map}

The loop paths of the cool coronal material in the \heI data set---those that exhibit redshifts greater than 50 km/sec---are manually traced using the {\sc Crispex} analysis tool \citep{vissers2012}.  {\sc Crispex} facilitates easy browsing of large spectral data cubes and has a built-in utility for determining loop paths for curvilinear structures.  While tracing individual loop paths, {\sc Crispex} permits scanning back and forth in spectral space to locate material exhibiting acceleration along their paths, which is critical for this application. 

Prior to using {\sc Crispex}, we restructured the FIRS data cube to an amenable format and additionally normalized each intensity spectrum by its local continuum intensity and by the median spectrum of the full field-of-view (see median profile in Figure~\ref{fig:firs_obs}).  This roughly removes the contribution of the blended telluric and photospheric features that otherwise would hinder the tracing procedure.  Importantly, the image shifts and drift correction described in Section~\ref{sec:obs_firs} is applied prior to tracing the loop paths. 

In total, 23 individual loops are identified in the \heI data as shown in Figure~\ref{fig:firs_traces}.  In the associated animation of panel (c), the flows do not appear to be uniform streams, but rather intermittent yet persistent downflows with variable column density as represented by EUV contrast.  The partial loop segments observed during the FIRS scan (\textit{e.g.}, \#8,11,17,14 in Figure~\ref{fig:firs_traces}) are likely short-duration flows consistent with localized blobs of down-flowing material as are often observed in filament drainage events and coronal rain.  

The degree of \heI absorption for each path varies substantially, with the structure consisting of loops \#4 and \#5 exhibiting much greater absorption than any of the other loops.  The \heI line depths of the down-flowing material in loops \#4 and \#5 are approximately 20\% of the continuum intensity, while the others are below 10\%.  Loops \#4 and \#5 also show more pronounced EUV contrast as well as more continuous flow relative to the other loops, making this structure easy to uniquely identify in the SDO/AIA images as well as STEREO/EUVI-A images below.  In addition, the \heI intensity and polarized spectra (discussed in Section~\ref{sec:pol_analysis}) are the strongest within this feature.  We therefore focus the remainder of our analysis on this coronal structure, and in particular on loop \#5; although, the results are equivalent for loop \#4.


\subsection{Stereoscopic triangulation of the Coronal Loop}\label{sec:stereo_recon}

We reconstruct the full three dimensional topology of loop \#5 in Figure~\ref{fig:firs_traces} by applying stereoscopic triangulation of its path trajectory, which has been aligned with the SDO/AIA 193 \mbox{\AA} observations, with cotemporal 195 \mbox{\AA} observations from STEREO/SECCHI/EUVI-A.  This process is illustrated in Figure~\ref{fig:stereo_fig}.  The spectral passbands of the 193 \mbox{\AA} and 195 \mbox{\AA} observations have similar temperature response.  Within the EUVI-A 195 \mbox{\AA} image acquired at 16:50:31 UTC, the observed active region is observed off-limb, and the down-flowing coronal loop is easily identified by its \edit1{comparatively stronger EUV extinction and temporal correspondence with the observed FIRS loop}. We have very high confidence that the structure observed with EUVI-A corresponds to the same structure comprised of loops \#4 and \#5 in Figure~\ref{fig:firs_traces}.  For the reconstruction, we cross-analyze this EUVI-A image with the SDO/AIA image acquired at 16:50:43 UTC, which best accounts for the 18.8 second light travel time difference between the Sun and the two spacecrafts on this date. 

The triangulation methods developed for the STEREO mission \citep{aschwanden2008,aschwanden2012} are applied to this loop using the AIA and EUVI-A images and header information (see Figure~\ref{fig:stereo_fig}).  This includes transforming the images and path coordinates into a common epipolar geometry in which the horizontal image axis aligns with the plane defined by the two spacecraft (see panels a and b).  We achieve this in much the same way as the SSWIDL procedure \textit{euvi\_stereopair}, and select to rescale the EUVI-A data to the projected pixel size of the higher resolution AIA images before feature triangulation.  Image centers and rotation angles are carefully verified throughout.  We then manually trace the outer edges of the $\sim 7''$ wide extinction feature in the EUVI-A image.  The points defining these edges are then cross-triangulated against the path trajectories from the \heI data set in the transformed epipolar geometry using the formulation of \cite{aschwanden2012}.  The result is the triangulated height of each point along loop \#5 as well as an error estimate that is consistent with the feature width in the EUVI-A image.  

As exhibited in Figure~\ref{fig:stereo_fig}, the FIRS \heI observations sample heights along loop \#5 from $\sim 3$ to 70 Mm above the solar surface.  The triangulated height error varies smoothly from $\sim \pm 2$ Mm to $\pm 3 Mm$ from the bottom to the top of the loop. The primary axis of the feature seen in panel (d) of Figure~\ref{fig:stereo_fig} is inclined with respect to the solar limb, and curls upwards, roughly in the solar north direction.  In Figure~\ref{fig:firs_traces} the loop extends well outside of the FIRS field-of-view, and extends up towards the location of the ``primary condensation" in Figure~\ref{fig:aia_images} where the coronal cloud filament (or spider) forms.  We associate the region of darker but diffuse material at heights above 0.1 $R_{Sun}$ in the EUVI-A image (Figure~\ref{fig:stereo_fig} panel (d)) with the formation of the coronal cloud filament.  


\section{Analysis}\label{sec:pol_analysis}

Having selected and triangulated the three dimensional topology of the coronal loop (\textit{i.e.}, loop \#5), we now proceed to analyze the \heI polarization signatures to constrain and derive the vector magnetic field structure of the loop. 


\begin{figure}
\centering
\includegraphics[width=0.45\textwidth]{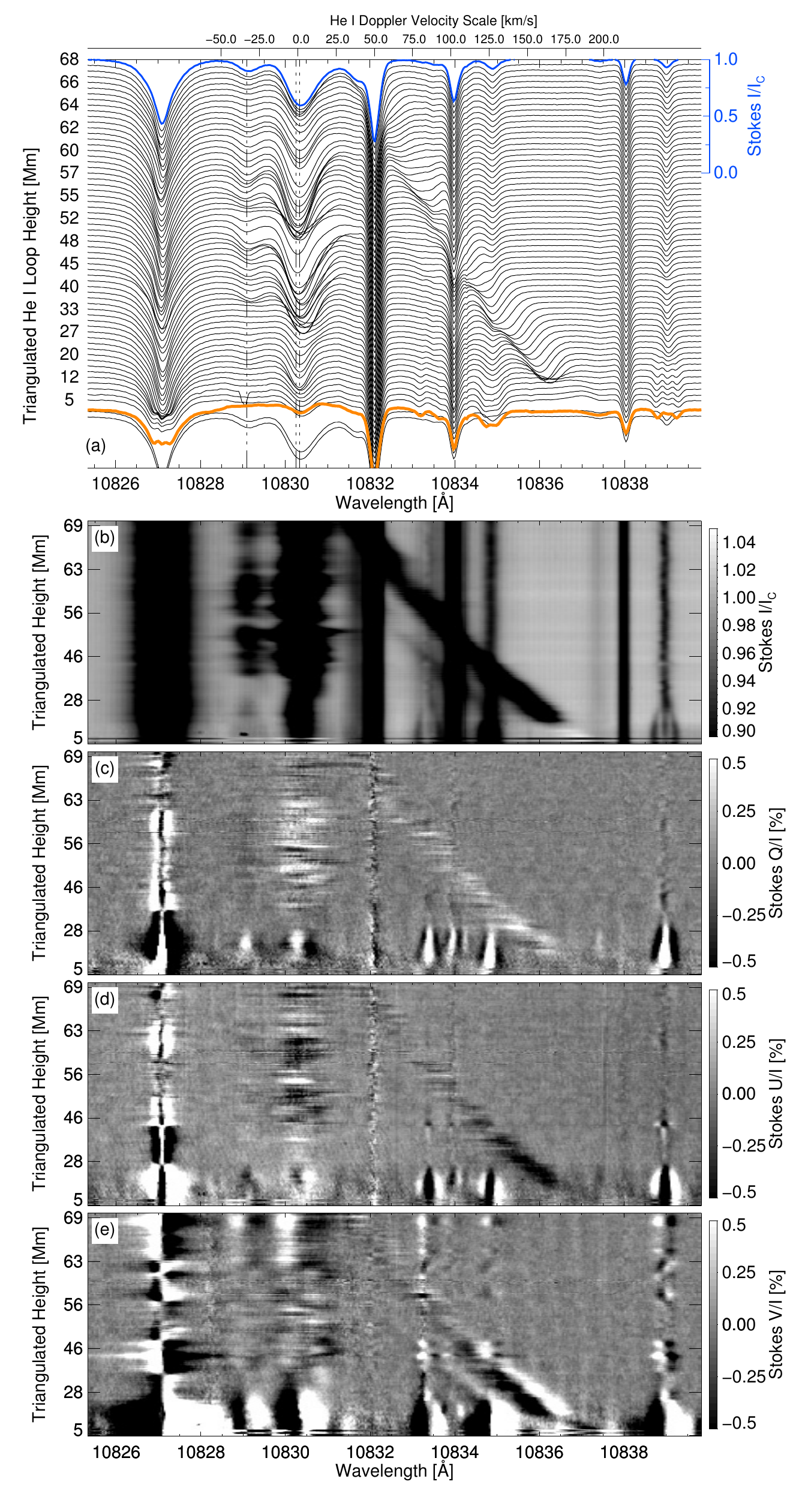}
\caption{Stokes spectra along the \heI coronal loop \#5 showing the acceleration of material along its path.  \textit(a) Stack plot of observed intensity spectra.  Heights given along the y-axis refer to physical heights of the fast \heI material above the solar surface as determined from a stereoscopic triangulation of the loop. Note that the sampling in height is not uniform.  Dashed-dotted lines mark the rest wavelengths of the \heI triplet. \textit(b-e) \heI Stokes spectra (displayed in image format) as extracted along the coronal loop path.}
\label{fig:rain_pol_spec}
\end{figure}

\subsection{\heI Polarization Signatures}\label{sec:heI_pol_sig}

The polarized (Stokes) spectra along loop \#5 are displayed as a function of triangulated height in Figure~\ref{fig:rain_pol_spec}.  These spectra are extracted directly from the DST/FIRS data cube by binning an area of 3 x 3 pixels around each point along the traced loop.  Binning further decreases the continuum noise level in Stokes (Q,U,V) to $(5.2,5.1,5.1) \times 10^{-4}$, in units of the incident intensity.  

At solar temperatures, the \heI 10830 \mbox{\AA} triplet ($1s2s{\ }^{3}S_{1} - 1s2p{\ }^{3}P_{2,1,0}$) consists of two resolved spectral features--a weaker `blue' line from the $2s{\ }^{3}S_{1} - 2p{\ }^{3}P_{0}$ transition, and a `red' line, which is a blend of the $2s{\ }^{3}S_{1} - 2p{\ }^{3}P_{1,2}$ transitions.  In addition, the spectra along loop \#5 contain two, well-separated, velocity components of \heI.  The  prominently red-shifted component, which originates from the coronal loop, exhibits Doppler velocities accelerating from $\sim 30$ km/s at the top of the observed loop (heights $\sim 68$ Mm) to $\sim 190$ km/s near the bottom (heights $\sim 4$ Mm).  As the material accelerates, the redshifted \heI line profiles traverse a number of other spectral features, as identified in Table~\ref{tbl:other_lines}. The median \heI line depth (\textit{i.e.}, $1 - I_{core}/I_{cont}$) in the fast component is 0.135 (maximum of 0.235), while the median equivalent width is 97.7 m\mbox{\AA} (maximum of 173 m\mbox{\AA}).

\begin{deluxetable}{llcc}
\tablecaption{Spectral lines analyzed in DST/FIRS spectral window \label{tbl:other_lines}}
\tablecolumns{4}
\tablenum{1}
\tablehead{
\colhead{Wavelength\tablenotemark{a} (\mbox{\AA})} &
\colhead{Line} &
\colhead{Log gf}  &
\colhead{Land\'e g$_{eff}$} 
}
\startdata
10827.0877 & Si \textsc{i}       & \phantom{-}0.185  & 1.5\phantom{0}   \\
10829.0911 & \heI (Tr1)          &           -0.745  & 2.0\phantom{0}   \\ 
10830.2501 & \heI (Tr2)          &           -0.268  & 1.75             \\ 
10830.3398 & \heI (Tr3)          &           -0.047  & 1.25             \\ 
10831.626  & H$_{2}$0 (telluric) &                   &                  \\
10832.108  & H$_{2}$0 (telluric) &                   &                  \\
10833.382  & Ca \textsc{i}       &           -0.515  & 1.0\phantom{0}   \\ 
10833.981  & H$_{2}$0 (telluric) &                   &                  \\
10834.848  & Na \textsc{i}       &           -1.800  & 1.03             \\
10834.848  & Na \textsc{i}       &           -0.500  & 1.07             \\ 
10834.907  & Na \textsc{i}       &           -0.650  & 0.90             \\
10838.034  & H$_{2}$0 (telluric) &                   &                  \\
10838.970  & Ca \textsc{i}       &           -0.033  & 1.5\phantom{0}   \\
\enddata
\tablenotetext{a}{Air wavelengths for atomic data other than \heI are sourced from the Kurucz databases \url{http://kurucz.harvard.edu/} \citep{kurucz1995}.  \heI data are from \cite{asensio_ramos2008}, while telluric line wavelengths are from \cite{breckinridge1973} and \cite{kuckein2012}.}
\end{deluxetable}

The loop's \heI polarized signatures show a large degree of spatial continuity (Figure~\ref{fig:rain_pol_spec}); that is, the dominant Q, U, and V profile shapes and sign are consistent along the loop.  Some variance is apparent, but we suspect it is introduced by terrestrial atmospheric seeing, and thereby discount the possibility that it is solar in origin.  The linearly polarized, normalized, Stokes Q/I and U/I signals in the \heI `red' line are weak (less than $0.5\%$) and single-lobed; we do not see evidence for a three-lobe shape indicative of the transverse Zeeman effect, as can be seen in the other solar lines within the sunspot (\textit{i.e.}, the spectral profiles associated with loop heights below $\sim28$ Mm).  The \heI `blue' transition does not show a signal above the noise level along the loop; however, within the other `slow' \heI velocity component, that associated with the lower lying chromospheric loops outside of the sunspot, the sign of Stokes Q in the \heI `blue' transition is opposite that of the blended `red' transitions, which, due to the polarizability of the \heI triplet, lends credence to the quality of the polarization crosstalk calibration (see \cite{schad2013}).  For both velocity components, the Stokes V profiles are Zeeman in nature, and, within the coronal loop, are evident along the entire loop and generally strengthen at lower heights and closer proximity to the sunspot umbra. 

We can interpret the loop's polarized spectra as the result of the joint influence of anisotropic radiative pumping, the Hanle effect, and the Zeeman effect \citep{trujillo_bueno_2002, asensio_ramos2008}.  Anistropic illumination of the \heI coronal material by the photospheric radiation field introduces atomic-level polarization, \textit{i.e.}, population imbalances and quantum coherences between the sublevels of degenerate atomic levels, which subsequently manifests as polarized spectral lines.  The presence of an inclined (relative to the radiation symmetry axis) magnetic field modifies the atomic-level polarization via the Hanle Effect for field strengths up to a critical field strength, after which the Hanle effect saturates and then no longer modifies the atomic-level polarization.  This regime is known as the saturated Hanle regime. For the \heI triplet lines, the critical field strength is 0.77 Gauss \citep{asensio_ramos2008}.  No significant Hanle modification occurs for field strengths approximately ten times this value, $\sim 8$ Gauss.  The single-lobed character of the observed loop's linear polarization (Q \& U) are consistent with the presence of atomic-level polarization in absence of a strong transverse magnetic field, \textit{i.e.}, one strong enough to induce large transverse Zeeman signals ($B_{trans} \gtrsim 250 G$).  Meanwhile, the measureable Stokes V signal (\textit{i.e.,} $>5.1 \times 10^{-4}$ $I_{c}$) along the loop places the field strengths to be above critical field strength, meaning the Stokes Q/U spectra are consistent with atomic-level polarization within the saturated Hanle regime. 

\begin{figure}
\centering
\includegraphics[width=0.45\textwidth]{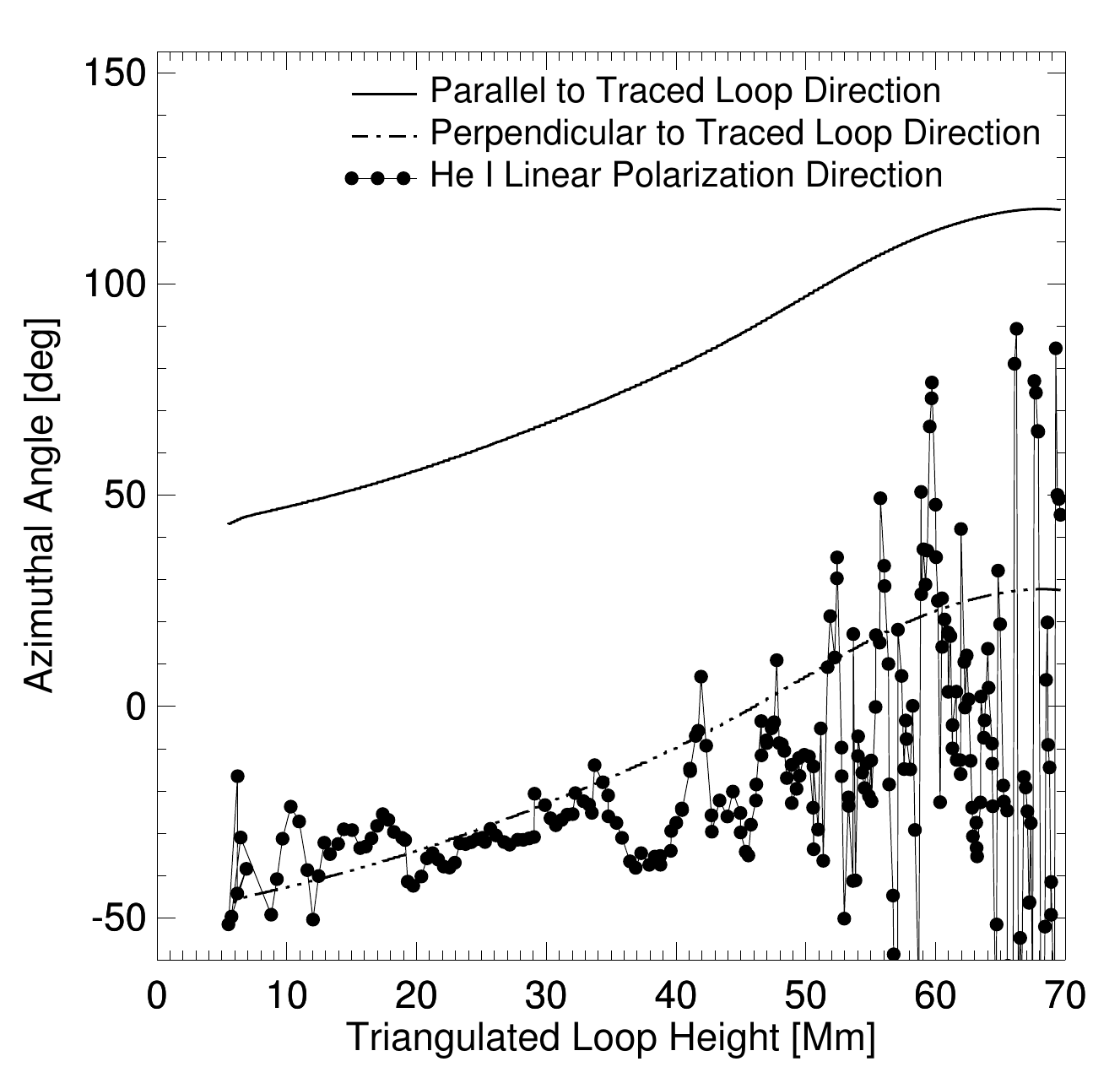}
\caption{Variation of the \heI red transition linear polarization angle projected on the plane of the sky (\textit{i.e.}, the azimuthal angle) as a function of the reconstructed (triangulated) coronal loop height.  For comparison, the angular orientation of traced coronal loop (solid line) is shown, alongside its orthogonal direction (dash dotted line).}
\label{fig:lin_pol_dir}
\end{figure}

In the saturated Hanle regime, population imbalances for a simple two-level atomic transition with $J_{lower} = 0$ (total angular momentum) and $J_{upper} = 1$ lead to linear polarization oriented either parallel or perpendicular to the transverse (relative to observer's line-of-sight) component of the magnetic field.  The action of lower-level polarization in a multi-level atomic system is able to alter this correspondence; however, under most observing conditions for \heI, the generalization above is valid, as discussed by \cite{schad2013}.  For an upper-level polarization dominated transition, like the \heI `red' transitions, the linear polarization direction is expected to be parallel to B if the inclination angle, defined as the angle between the local solar vertical and the magnetic field vector, is greater than the Van Vleck angle, $\theta_{V} = 54.74^{\circ}$, and less than $(\pi - \theta_{V}$).  Otherwise, it is perpendicular to B.  

Figure~\ref{fig:lin_pol_dir} shows the linear polarization angular direction (azimth; in the plane of the sky) as compared to two curves representing the angular direction of the loop traced in Section~\ref{sec:trace_map}. We determine the direction of linear polarization along the coronal loop as follows:
\begin{equation}
\chi = 
\frac{1}{2} \arctan \left ( 
\frac{\sum_{\Delta\lambda = - 193{\ }m\mbox{\AA}}^{+193{\ }m\mbox{\AA}} {\ }U(\lambda_{c} + \Delta\lambda)}
{\sum_{\Delta\lambda = - 193{\ }m\mbox{\AA}}^{+193{\ }m\mbox{\AA}} {\ }Q(\lambda_{c} + \Delta\lambda) } \right )
\end{equation}
where $\chi$ is the angle of linear polarization referenced to the Stokes +Q reference axis, \textit{i.e.}, parallel to the solar equator.  This value has a $180^{\circ}$ ambiguity. $\lambda_c$ refers to the line center position of the `red' \heI transitions, which is found via spectral profiles fits described below in Section~\ref{sec:me_inversions}.  Our stereoscopic triangulation (Figure~\ref{fig:stereo_fig}) determined that the inclination of the traced loop is always less than $\theta_{V}$, meaning that if the magnetic field direction and the loop path correspond, the linear polarization direction (in the saturated Hanle regime) is expected to be oriented perpendicular to the traced loop.  We do find such an agreement despite the level of measurement noise, with a very convincing correspondence for heights between 10 and 35 Mm.  The level of scatter in the measured azimuth angle increases for heights above $\sim45$ Mm where the role of noise (likely due to more pronouced seeing polarization effects away from the adaptive optics isoplanatic patch) in the linear polarized spectra begins to dominate.  We conclude that our polarized spectra are consistent with a magnetic field aligned structure, as is generally expected, though not often proven. 

\begin{figure*}
\includegraphics[width=\textwidth]{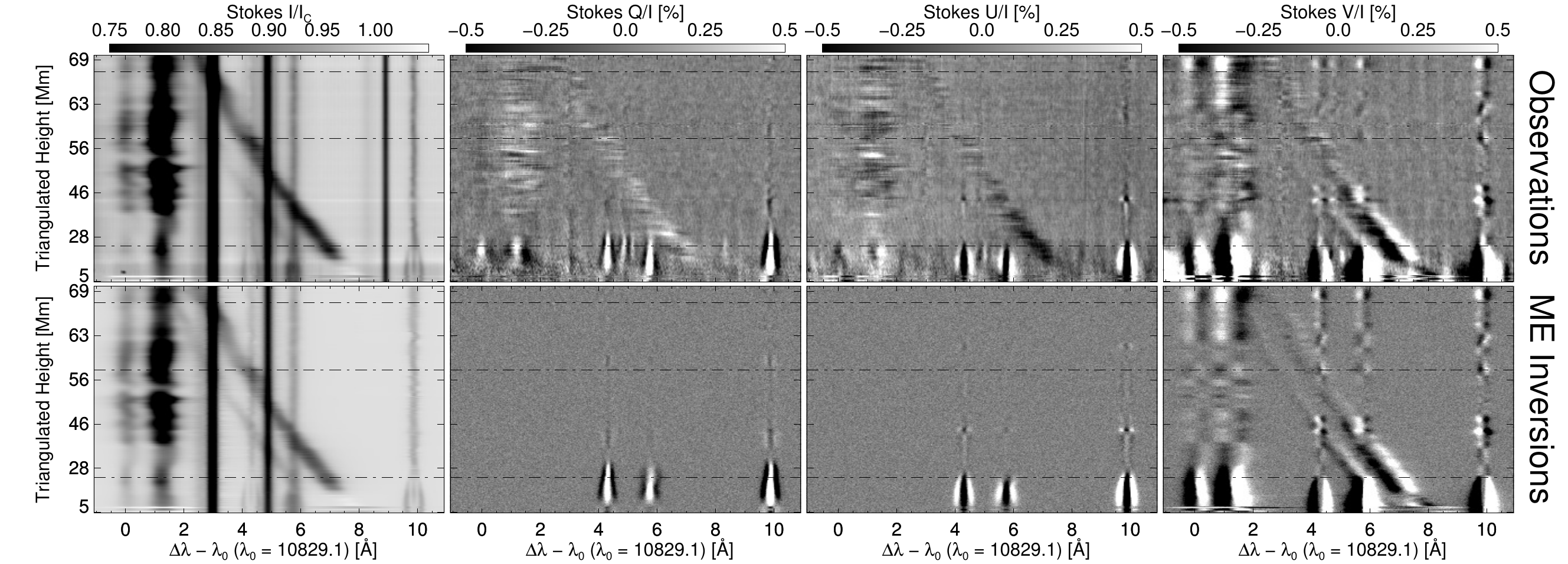} \\ 
\includegraphics[width=\textwidth]{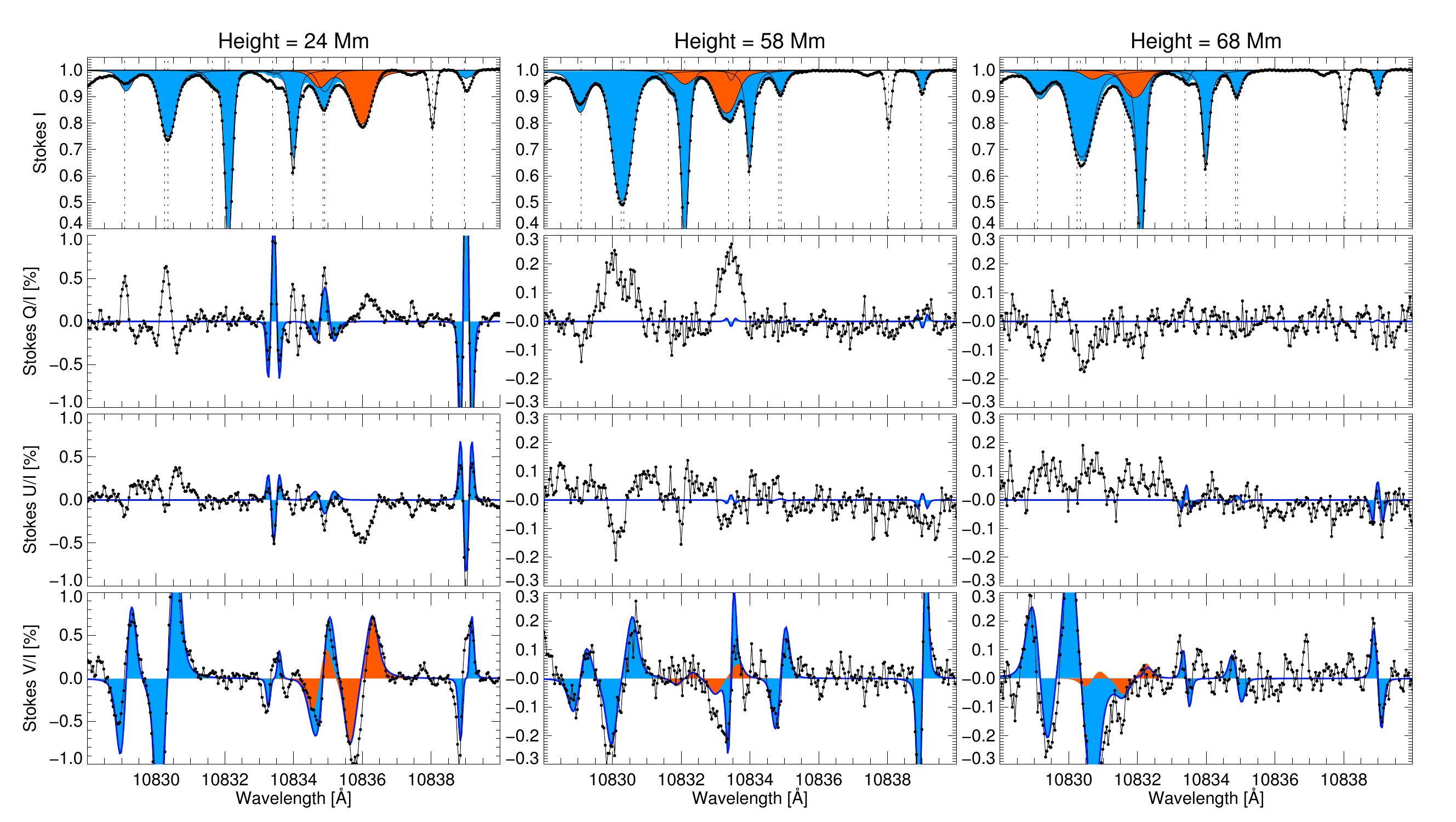} \\
\caption{Multi-line Milne-Eddington (ME) inversions of the spectral profiles along the coronal loop.  The full Stokes vector is inverted for each photospheric Zeeman-sensitive line, as discussed in the text, while only the Stokes I and V (not Q \& U) profiles are fit for the \heI spectra by using an ME model atmosphere of a strictly longitudinal magnetic field. \textit{(top)} A comparison of the observed and inverted Stokes spectra shown as a spectral image.  An artificial noise consistent with the observed noise is added to the inverted spectral image. \textit(bottom)  Individual line profiles and inversion results for three representative loop heights.  Observed profiles are given by black lines connecting black dots. Filled blue areas give the full inverted spectra, while the orange filled areas denote the \heI components of the fitted spectra.  For Stokes I, the other line components are also given by thin black lines. Vertical dashed lines indicate the rest wavelengths of the observed lines.}
\label{fig:me_inv_fits}
\end{figure*}


\subsection{Spectropolarimetric Inversions}\label{sec:inversions}

To infer the full magnetic field vector along the coronal loop, we perform spectropolarimetric inversions using the flexible code called \helix \citep[from the \underline{He}-\underline{L}ine \underline{I}nformation E\underline{x}tractor$^\textbf{+}$,][]{lagg2004,lagg2007a}.  \helix numerically optimizes a parametric radiative transfer model to best fit observed Stokes spectra, on a pixel-by-pixel basis.  Here, we employ two independent radiative transfer models.  First, a simple, computationally-efficient, Milne-Eddington (ME) model (Section~\ref{sec:me_inversions}) is applied, using multiple steps with careful constraints, to invert the \heI profiles, as well as all the associated line blends between 10829 and 10840 \mbox{\AA}.  The ME atmosphere used includes the Zeeman effect, but does not account for the atomic level polarization and Hanle effect signatures of \heI; thus, only the longitudinal flux density along the loop can be inferred by this model.  After successful ME inversions, we subtract the blended line signals from the fast \heI component, and then apply the ``Hanle slab'' model, which accounts for both the Hanle and Zeeman effects (Section~\ref{sec:hanle_inversions}), to only the fast \heI spectra to derive the full magnetic field vector along the coronal.  Both the ME and Hanle slab models are available in \helix; though, they cannot currently be used simultaneously for one profile inversion. 


\subsubsection{Multi-line Milne-Eddington (ME) Inversion Strategy}\label{sec:me_inversions}

The influence of the background spectral lines that the coronal loop's \heI signal traverses as the material accelerates must be carefully managed during our inversion analysis.  At lower heights, in particular, the loop's \heI Stokes V profiles become blended with the photospheric Zeeman-sensitive lines of Ca I (10833.38 \mbox{\AA}) and Na I (three transitions near 10834.85 \mbox{\AA}).  The \heI linear polarized signals (Q \& U) are only minimally affected by the background lines, as the majority of the loop is located above relatively quiet regions of the photosphere.  

To decouple the \heI signals from the background photospheric lines and telluric lines, we developed a multi-step procedure that results in a well-constrained multi-line inversion of each observed profile.  Each solar spectral line is ultimately modeled using a Milne-Eddington (ME) atmosphere (see, e.g., \citep{landi_2004}) using the following free parameters: magnetic field strength (B), inclination ($\gamma$), azimuth ($\phi$), line strength ($\eta_{0}$), Doppler width ($\Delta\lambda_{D}$), a damping parameter, line-of-sight velocity ($v_{LOS}$), the source function at $\tau = 0$, its gradient, and the filling factor (f).  Throughout our analysis, we assume a unity filling factor, meaning the magnetic field uniformly fills the resolution element. As a result, we use the terms `magnetic field' and `magnetic flux density' interchangeably.  For the \heI 10830 \mbox{\AA} line, the polynomial approximation of the Paschen-Back Effect is included in the magnetic line splitting calculation \citep{socas_navarro2005}.  We also allow the gradient of the \heI source function to be either positive or negative, so to force the fitting of emission profiles.  Additional Voigt profiles are fit to the the telluric spectral features.  We use the {\sc PIKAIA} genetic function optimization algorithm \citep{charbonneau1995}, as encoded in \helix, throughout, using a suitably large number of {\sc PIKAIA} generations.  The overall procedure is summarized in the following steps:  

\begin{enumerate}[noitemsep]
\item \textit{Fits for telluric lines near 10832 \mbox{\AA}}:  Using the entire DST/FIRS data cube, we use \helix to fit a preliminary model comprised of a single ME component describing the `slow' (not the coronal loop) \heI velocity component and voigt profiles describing the two telluric features near 10832 \mbox{\AA}.  The fit only uses the intensity spectra (Stokes I).  Owing to the fact that the `slow' component of \heI is typically well resolved in relation to the telluric features, in addition to the limited presence of the cool coronal material, the median profile fit to the telluric lines for each slit position sufficiently constrains the telluric profile shapes for each step position. \par 
\item \textit{Determine median background spectra between 10832 and 10838 \mbox{\AA}}:  To isolate the background spectra between 10832 \mbox{\AA} and 10838 \mbox{\AA}, we first divide out the telluric lines derived in Step 1.  All intensity spectra are then sorted by continuum intensity.  A median intensity profile is determined for specified ranges of continuum intensity, thereby creating a library of median profiles as a function of continuum intensity.  Since the photospheric magnetic field strength is strongly correlated with intensity (see, e.g., \cite{schad2014}), these median profiles approximately account for the magnetic-field-dependent Zeeman-broadening of each background solar line.  Meanwhile, they do not contain any significant contribution from the coronal loop's \heI spectra.  
\item \textit{Separation of background spectra from \heI spectra}: Beginning with the extracted spectra along the coronal loop (see Figure~\ref{fig:rain_pol_spec}), we divide out the telluric lines near 10832 \mbox{\AA} as derived in Step 1.  We then compile the background spectra between 10832 \mbox{\AA} and 10838 \mbox{\AA} according to the observed continuum intensity by using our library of median profiles calculated in Step 2.  Finally, we divide the original intensity spectra by the derived background intensity spectra.  This approximately separates the Stokes I contribution of the background lines from the \heI profiles, with residuals peaking at $\sim 2\%$ of the continuum intensity, \textit{i.e.,} significantly less than the median line depth of loop's \heI signal.  The results are three separate versions of the spectra along the loop:  (1) as observed, (2) the approximated background spectra lines, and (3) the isolated \heI spectral lines.  Note that only the Stokes I profiles are separated from each other in this process.  We address the blended polarized spectra (Q,U,\&V) via the fits in Steps 5 and 6. \par
\item  \textit{Fit Isolated \heI Intensity Spectra:} The isolated \heI intensity spectra (Stokes I) are fit using a two component ME atmosphere, one for each velocity component.  We do not consider the polarized spectra, and thus only derive the thermodynamic parameters of the \heI intensity spectra, which we record for use in Step 6. \par 
\item \textit{Fit spectral lines in background spectra}:  The background lines of Ca \textsc{i} (10833.382) and Na \textsc{i} (near 10834.85 \mbox{\AA}) are fit with ME models that include a magnetic field (unlike Step 4).  However, the \heI polarized signal is not removed in the steps above, and may influence the fits to these spectral lines.  To address this issue, we include a fit for the unblended Ca \textsc{i} line near 10838.97 \mbox{\AA}, and force the magnetic field parameters of the blended Ca \textsc{i} and Na \textsc{i} lines to take on its magnetic field values as the profiles are inverted.  This is a justified approximation due to the similarly weak line strength of these lines.  The result of this step is the fitted parameters of the ME model for each background spectral line, in addition to voigt line fits to the telluric line at 10833.981\mbox{\AA}.  \par 
\item \textit{Combined fits}: Finally, we invert the original Stokes spectra extracted along the loop with ME model atmospheres describing each spectra line, including all background solar lines and the two \heI velocity components.  This time we include a magnetic field for the \heI components, but force it to be longitudinal so to fit only the Zeeman effect signals observed in Stokes V.  During the fit, the velocity, Doppler width, and source function gradient parameters for each line other than the Ca \textsc{i} lines are allowed to vary by up to 10 percent relative to the fits derived in Steps 4 and 5.  The Na \textsc{i} magnetic field intensity is allowed to vary also by up to 10 percent.  All other parameters are held fixed.  
\end{enumerate}

Figure~\ref{fig:me_inv_fits} shows the results from this rather involved Milne-Eddington inversion.  By inspection of the fits and the resulting parameters, we found the procedure performs well.  The various spectral line signals are well recovered, even when blended with the highly redshifted \heI profiles.  Note in particular the fits to Stokes V in the line profiles displayed in the figure.  For loop heights of 24, 58, and 68 Mm, the Stokes V signal of the redshifted component of \heI is blended with the Na \textsc{i}, Ca \textsc{i} and \heI, respectively; yet, the relative large strength of those background V signals do not lead to anomalously large magnetic fields in the fast \heI component.  Furthermore, note that the linear polarized signals of the fast \heI component are preserved during the fitting.  In particular, for the spectrum marked with a height of 58 Mm, the blended Ca \textsc{i} line does not attempt to fit for the added linear signal.  Instead, it is constrained by the fit to the Ca \textsc{i} at 10838.97 \mbox{\AA}.  Overall, we are confident that our approach achieves satisfactory multi-line fits to these spectra.


\subsubsection{\heI Milne Eddington Results for Longitudinal Field}\label{sec:me_results}

\begin{figure}
\centering
\includegraphics[width=0.5\textwidth]{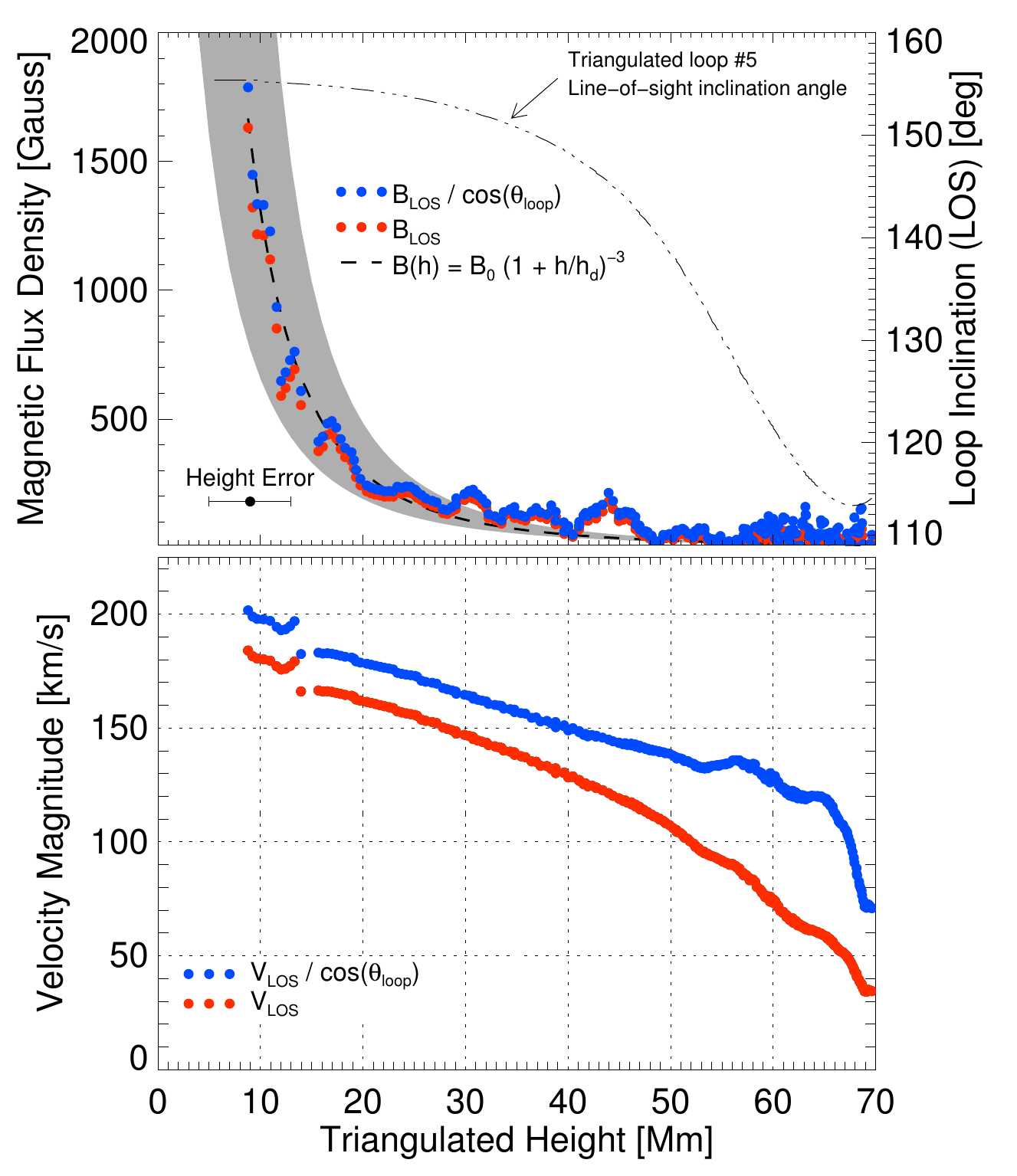}
\caption{Results from the Milne-Eddington analysis showing the variation of the \heI 10830 \mbox{\AA} magnetic flux density (\textit{top}) and velocity (\textit{bottom}) as a function of the triangulated height along the observed coronal loop.  For both quantities, the ME analysis only derives the longitudinal (line-of-sight; LOS) component.  Under the assumption that the magnetic field and mass flow is directed along the physical orientation of the traced loop, the corrected loop-aligned flux density and velocity are given as shown.  The dashed line is a fit of the observations to the dipole potential field approximation in the text, while the gray area denotes the range of fits consistent with the observations and estimated height error.  Derived values for $B_{0}$ are much larger than expected.  See text for details.}
\label{fig:me_inv_results}
\end{figure}

Figure~\ref{fig:me_inv_results} shows the variation of the \heI 10830 \mbox{\AA} magnetic flux density (\textit{top}) and velocity (\textit{bottom}), inferred  by the multi-line Milne-Eddington inversions, as a function of the triangulated height along the observed coronal loop.  Note that our ME model of \heI only infers the line-of-sight components of the magnetic flux density and velocity.  We also provide data (blue points) for the total magnetic flux density and total velocity under the assumption that the field and flow aligns with physical orientation of the stereoscopically triangulated loop. The relative difference between the longitudinal and total field/flow components range between $\sim13$ and $\sim50\%$ from lower to upper heights due to the projection of the loop relative to the line of sight.  Note also the estimated height error of +/- 4 Mm, which is a combination of the triangulation error and our estimate for the coregistration error. 

The loop-aligned total velocity of the cool \heI material peaks nears 200 km/sec at a height of $8 \pm 4$ Mm.  The terminal velocity of free-falling material in the corona, initially at rest, (\textit{i.e.}, $\sqrt{2gh}$ where g = 274 km/sec) is greater than 200 km/sec for heights larger than 73 Mm.  Based on our stereo observations of the loop heights, these observations are consistent with free-falling coronal rain material.

The derived magnetic flux density decreases rapidly from the lower atmosphere to the upper corona, as is generally expected.  Here we compare the observed decrease with that expected from a simple dipole potential field model.  As shown by \cite{aschwanden2005corona}, an analytic approximation of the dependence of the magnetic field strength (B) on vertical height (h) above the photosphere can be written as
\begin{equation}
B(h) = B_{0} \left ( 1 + \frac{h}{h_{d}} \right ) ^{-3},
\label{eq:dipole_approx}
\end{equation}
where $h_{d}$ refers to the \textit{dipole depth}, which is defined as the location where the magnetic field becomes singular. At $h = 0$ (the solar surface), the magnetic field intensity is $B_{0}$.  We fit this functional form to our observations, and find a fair, qualitative correspondence ($B_{0} = 29380; h_{d} = 5.5 Mm;$ see dashed line in Figure~\ref{fig:me_inv_results}). \edit1{By subjecting the abscissa to scalar height errors of $\pm 4$ Mm, we derive outer bounds on the fit parameters to be $B_{0} = 5688 G; h_{D} = 9.5$ and $B_{0} =388890 G ; h_{D} = 2.4$.  In all cases, the inferred value for $B_{0}$ is above those expected in the photosphere, suggesting the loop field strengths in the corona are stronger than those given by the dipole potential field approximation.}


\subsubsection{\helix ``Hanle-slab'' Inversions}\label{sec:hanle_inversions}

\begin{figure*}
\centering
\includegraphics[width=\textwidth]{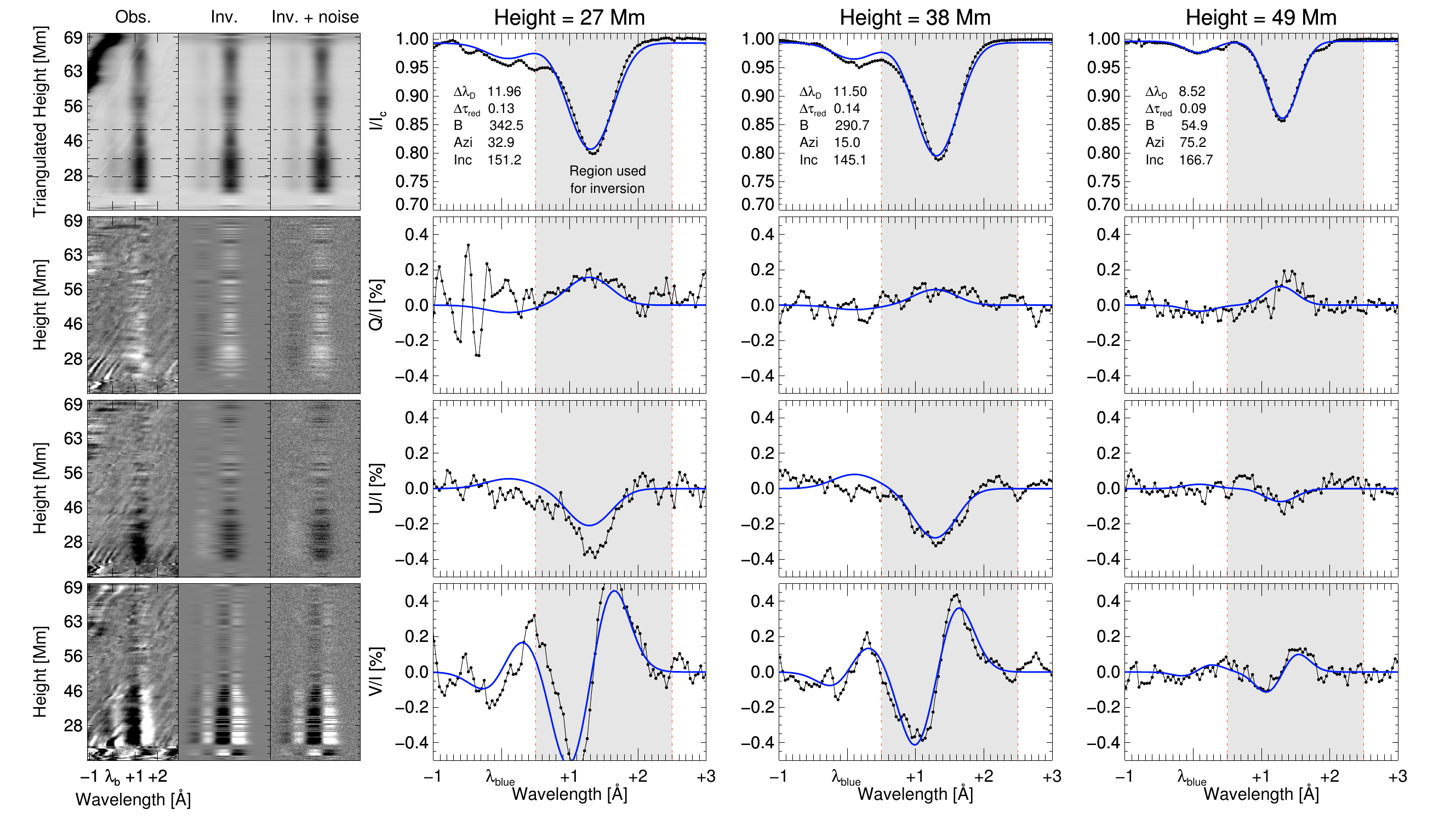}
\caption{Inversions of the \heI 10830 \mbox{\AA}  coronal loop Stokes spectra using the ``Hanle slab'' model of \helix . The image spectra in the left hand column display the observed profiles as a function of the triangulated height along the coronal loop.  The large Doppler shifts along the loop have been subtracted in order to align all loop spectra.  Alongside the observations in the first column, the inverted spectra are displayed both without (middle) and with (right) added artificial noise consistent with the observational noise.  Line fits to individual profiles are displayed in the three columns on the right for three heights along the loop as indicated.  Although the full inverted profiles are shown, only the region indicated by the light blue area is used during the inversion process in order to reduce residual errors due to line blends.}
\label{fig:hanle_inv_fits}
\end{figure*}

Building upon the results from the Milne-Eddington analysis above, we move now to the full Stokes analysis of the highly redshifted He I spectra along the coronal loop.  We apply the ``Hanle slab'' model of \helix, which is an equivalent implementation of the forward model described by \cite{asensio_ramos2008} and used in the associated {\sc Hazel} inversion code.  This radiative transfer model takes into account atomic level polarization induced by anisotropic radiation pumping, as well as the Hanle and Zeeman effects, under the assumption that the magneto-thermal properties are constant within a narrow slab of material suspended at a discrete height above the solar atmosphere.  The radiation anisotropy originates from the directional, and limb-darkened structure of the underlying photosphere, which is a function of the height above the solar atmosphere.  Higher order anisotropies induced by structures on the solar surface, e.g., sunspots, are not taken into account.  Up to eight free parameters are allowed to fit the He I Stokes profiles: the optical depth of the line ($\Delta\tau_{red}$), the line damping (a), the Doppler width ($\Delta\lambda_{D}$), Doppler velocity, the three components of the magnetic field vector, and the material height (h).  The filling factor is assumed to be unity, as for the ME analysis above.

The Hanle slab model is only applied to the highly redshifted He I component, and the line blends are not simultaneously fit due to the inability of \helix to fit both ME and Hanle slab models in parallel.   Instead, we fit the intensity spectra of isolated \heI spectra (see Section 4.2.1. Step 3) and its polarized spectra.  However, we subtract the polarized spectra of background line fits of section 4.2.1. to suppress the influence of line blends on the results.  Each \heI spectral profile is also shifted in wavelength according to the Doppler shift derived by the ME model, which eliminates the need to optimize a solution over the large range of Doppler velocities.  To further suppress the role of noise and blends, only the strong red transition of the \heI triplet is utilized to determine the fit parameters.
 
Optimal Hanle slab model solutions are found using two inversion steps.  The first step optimizes the fit to the Stokes I spectra using only the Doppler width and optical depth as free parameters.  The line damping parameter (a) is calculated directly from the Doppler width, using the Einstein coefficient of the transitions, which accounts for thermal damping effects only.  The height is fixed for each observed spectrum according to the stereoscopically triangulated heights.  The magnetic field intensity is held fixed at 100 G during this step.  In order to fit the \heI emission profiles for heights below 14 Mm, we permit the optical depth of the slab to take on negative values\footnote{Alternatively, we considered fitting for a scalar enhancement of the source function, which is a functionality available in \textsc{Hazel}, but not \helix, for the modeling of emission profiles.  Tests did not find any appreciable differences for our application between this method and the \edit1{\textit{shortcut}} use of a negative optical depth\edit1{; although, we stress the use of an enhanced source function is the correct approach.}}.  A single solution for the two free parameter is found using the damped least-squares (\textit{i.e.,} \edit1{Levenberg-Marquardt}) method. 
 
The second inversion step determines the three parameters describing the vector magnetic field, with all other parameters held fixed.   Here we employ the {\sc Pikaia} algorithm for optimization.  The advantage of {\sc Pikaia} for our application it that its stochastic path to convergence is non-deterministic, which makes it very effective at locating ambiguous solutions when performing repeated fits.   Consequently, we repeatedly invert each Stokes spectra along the coronal loop 12 times using the {\sc Pikaia} algorithm to locate ambiguities. 
 
Figure~\ref{fig:hanle_inv_fits} shows the resulting Hanle slab fits to the highly redshifted \heI spectra of the coronal loop.  Despite the variance of the observed signals along the loop (likely due to terrestrial atmospheric seeing), the inversions do well to recover the primary aspects of the observed Stokes spectra, and also show that the presence of the observational noise makes it difficult to confidently identify linear polarized signals in the \heI blue transition. 


\subsubsection{Coronal Vector Magnetic Field Results}\label{sec:hanle_results}

\begin{figure*}
\centering
\includegraphics[width=\textwidth]{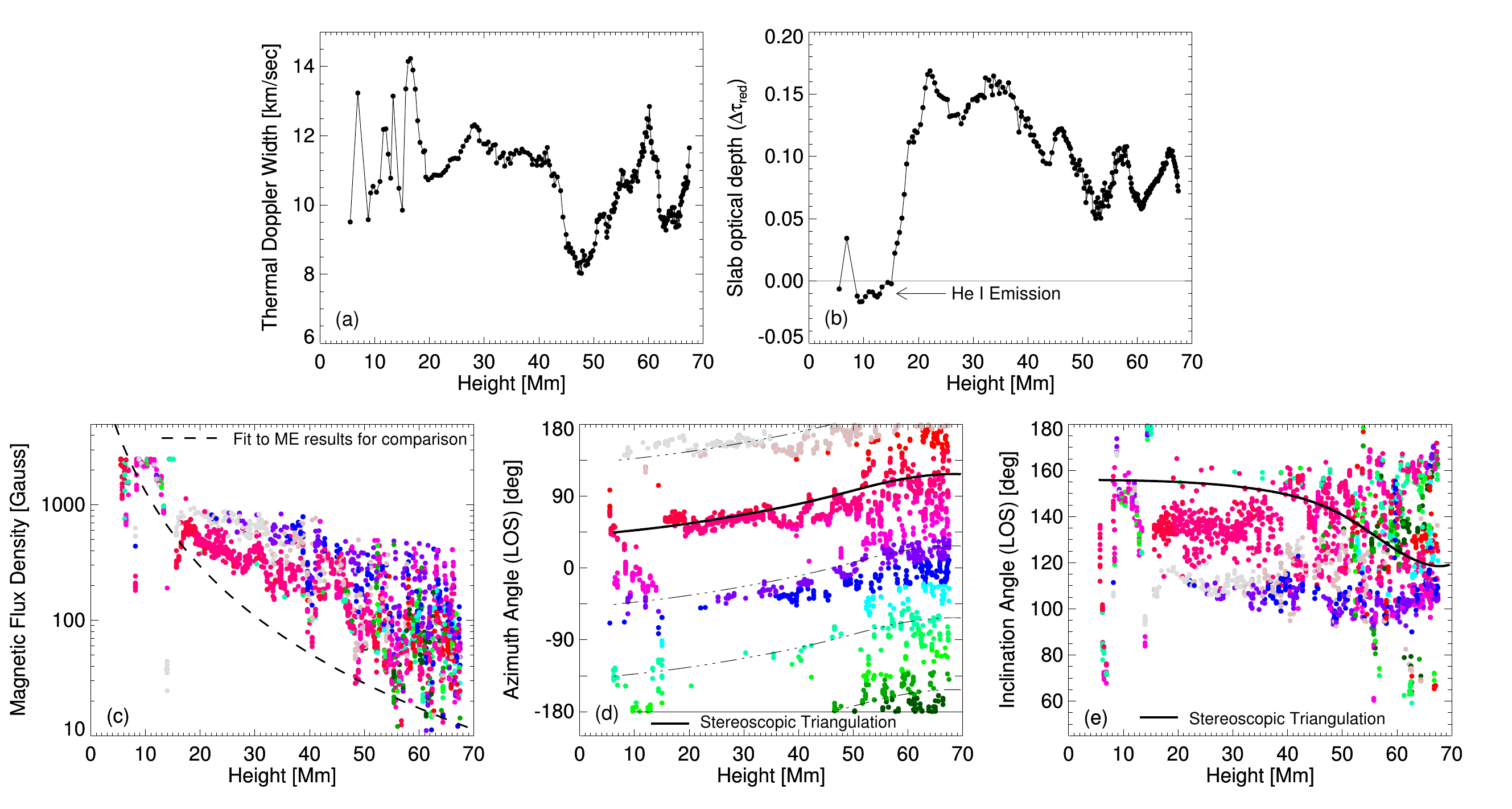}
\caption{Results from the Hanle slab model analysis showing, as a function of the triangulated height along the observed coronal loop, the variation of the (a) thermal Doppler width, (b) slab optical depth, (c) magnetic flux density, (d) magnetic field azimuth (here it is the orientation projected in the plane of the sky), and (e) magnetic field inclination angle measured relative to the line-of-sight. The dipole appoximation fit from the ME analysis (see Figure~\ref{fig:me_inv_results} and Equation~\ref{eq:dipole_approx}) is given for comparison in panel (c).  Panels (d) and (e) also show the measured orientation of the triangulated coronal loop (solid lines).  In panel (d), the orthogonal and anti-parallel azimuthal orientations, indicative of saturated Hanle effect ambiguities, are also shown.  Each unique inverted solution in panels b-e is colored according to its deviation from the traced azimuthal orientation (solid line); red data points indicate the solution with azimuths matching the traced loop.}
\label{fig:hanle_inv_results}
\end{figure*}

Figure~\ref{fig:hanle_inv_results} shows the results of the Hanle slab model inversions, including the Doppler width, the optical depth of the slab, and the three parameters of the vector magnetic field, for which coherent results are obtained for heights ranging between 14 and 50 Mm.  Outside of this range, the strength of the linear polarized signals is insufficient for reliable vector fits (see Figure 9).
 
Although the Hanle slab model inverts for the magnetic field vector in a coordinate frame local to the material, we transform the vector into a line-of-sight coordinate frame.  Thus, azimuthal angles are measured in the plane formed by the line-of-sight and the plane of the sky, and reference the Stokes +Q direction (i.e. aligned with the solar equator).  Azimuths with a value of zero are directed toward Solar East; values of +90 are directed towards Solar South.  Meanwhile, the inclination angle is measured between the line-of-sight and the magnetic field vector, with values of 0 being directed towards the observer.
 
The repeated inversions discussed in the previous section result in multiple identified, ambiguous solutions  (see panels c-e of Figure) , each of which sufficiently match the observed Stokes profiles.  As can be seen in panel (d), these solutions primarily fall along four curved lines, which denote the azimuthal orientation of the traced coronal loop and its orthogonal and antiparallel orientations.  The results are consistent with a magnetic field aligned with the traced coronal loop axis, as already suggested by the direction of linear polarization (i.e. Figure~\ref{fig:lin_pol_dir}).  The classes of solutions following the orthogonal paths are consistent with Van Vleck ambiguities, as discussed in Section~\ref{sec:heI_pol_sig}.  Fewer solutions lie along the antiparallel direction because the sign of Stokes V breaks the ambiguities for most spectra profiles.  The class of solutions that align with the coronal loop (colored in red) are those that most closely match the reconstructed inclinations and magnetic flux densities inferred by the ME analysis.
 
Yet, the inferred inclination angles are slightly more inclined ($\Delta\chi  \sim~20^{\circ}$) relative to the line-of-sight than the stereoscopic reconstruction values while the magnetic flux densities are slightly larger than in the Milne-Eddington analysis.  These variances are likely linked and related to the observational noise in the data.  As a result of the profiles being primarily formed in the saturated Hanle regime, the magnitude of linear polarization is insensitive to the magnetic field intensity but is a primary constraint on the field inclination relative to the solar vertical.  Due to the nearly vertical orientation of the loop as it connects to the solar photosphere, the sensitivity of the linear polarized signals to changes in inclination are weak.  The systematic differences of $\sim20^{\circ}$ between the inverted inclinations and the expected inclinations derived via stereoscopy are not unexpected due to the presence of observational noise.  As a result of the systematically more inclined fields, the inferred total magnetic flux density increases to preserve its fit for the longitudinal field strength constrained by Stokes V.
 
Still, as a whole, the Hanle slab inversions recover well the vector properties of the cooled coronal loop, and despite the slight offsets in the inferred inclinations, lend further credence to the conclusion that the three dimensional coronal loop morphology is indeed aligned with the vector magnetic field structure along the loop.


\section{Discussion}\label{sec:discussion}


\subsection{Multi-component high-velocity \heI phenomena}

Multiple velocity components of the He I triplet have previously been reported by \cite{schmidt2000}, \cite{lagg2007b}, and \cite{sasso2011}. The high velocity He I material described in this paper exhibits redshifts up to 185 km sec$^{-1}$, considerably larger than the previously greatest reported redshift of 100 km sec$^{-1}$ for He I triplet observation of an active filament \citep{sasso2011}.  A second supersonic down-flowing component appears to very commonly appear in He I spectra, with LOS velocities between 10 and 40 km sec$^{-1}$ \citep{aznar_cuadrado2007, gonzalez_manrique2016}.   \cite{penn2000} observed strongly blue-shifted He I profiles of an erupting active region filament, with upward velocities between 200 and 300 km sec$^{-1}$.
 
The strong \heI downflows we observe closely resemble the supersonic downflows observed by \cite{kleint2014} in transition region (TR) spectral lines above sunspots. As in our case, \citeauthor{kleint2014} associated the strong flows with coronal rain produced within loops undergoing a thermal instability, and reported intensity brightenings at the loop footpoints consistent with a temperature and density increase in the TR, most likely the equivalent of the shock we observe at the footpoint of the \heI coronal loop (see section~\ref{sec:obs_firs}).  Furthermore, the velocities that we measure are consistent with gravitational free fall from heights above 70 Mm, consistent with our stereoscopic observations.  Unlike \citeauthor{kleint2014}, we do not see evidence for velocities greater than free-fall from a stationary state.

Our observations show the draining loop's footpoint is rooted near the sunspot's umbra-penumbral boundary near a small light bridge (see Figure~\ref{fig:firs_traces} panel (a)), which is similar to the \citeauthor{kleint2014} observations (see also \cite{ahn2014}). \cite{chitta2016} reported downflows of more than 100 km sec$^{-1}$ rooted within a sunspot umbra, which due to the long duration of the flows ($\sim 22$ min) and estimates for the mass supply rates of the draining loops, were associated with a possible siphon flow instead of coronal rain.  However, the coronal rain event studied here, also shows persistent downflows into the sunspot lasting at least 30 minutes, suggesting the mass supply rates are sufficient for persistent drainage over tens of minutes.  We speculate that the apparent expansion rate of the coronal volume experiencing the drainage (see Figure~\ref{fig:firs_traces} panel (a)) is much larger than the uniform coronal loop cross-section assumed by \citeauthor{chitta2016}.  


\subsection{Coronal magnetic field measurements}

To our knowledge, these observations provide the first vector magnetic field inferences along an individual, resolved coronal loop structure extending up to heights of 70 Mm ($0.1$ $R_{sun}$), and the first polarimetric measurements of coronal loops located on disk.  Recent off-limb work has advanced high-dynamic range observations of weak forbidden coronal emission lines in the infrared for diagnosing spatially resolved coronal fields \citep{lin2004, tomcyzk2008}; though, off-limb studies in optically-thin spectral lines must take into account line-of-sight integration effects \citep{liu_lin2008,dove2011}.  The disadvantage of on-disk observations is typically the inability to constrain the material height; however, here we solve this problem via stereoscopic observations.  

Our results compare favorably with the few previous measurements of coronal magnetic fields.  Seventy minute polarimetric integrations of the forbidden Fe \textsc{XIII} 10747 \mbox{\AA} coronal emission line by \cite{lin2004} inferred off-limb \textit{longitudinal} flux densities of 4 G at a projected height of $100''$ (72.5 Mm) above an active region.  Similarly, our measurements support weak (10s of Gauss) \textit{total} flux densities at heights of 70 Mm.  \cite{tomczyk2007} applied coronal seismology techniques to off-limb observations of the Fe \textsc{XIII} line (at projected heights above 35 Mm) and inferred average coronal magnetic field strengths between 8 and 26 G using the Alfv{\'e}n wave phase relation.  Thermal bremsstrahlung and gyroresonant radio emission can also be used for comparatively lower spatial resolution coronal magnetic field measurements; 15 GHz and 8 GHz measurements by \cite{brosius2006} inferred coronal field strength above a large sunspot of 1750 G and 960 G, at heights of 8 Mm and 12 Mm, respectively.  These values compare well with our values of 1980 G and 900 G at these heights. 

Yet, our observations are fundamentally different than earlier results in that we probe cool, neutral material of a thermally instable loop, and not mega-kelvin, fully-ionized coronal plasma.  We estimate upper limits of the \heI material's temperature, based on observed \heI Doppler widths (8 to 12 km sec$^{-1}$), to be between 15 and 35 kilo-Kelvin, in accordance with similar measurements using H$\alpha$ and the Ca \textsc{ii} 854.2 nm lines by \cite{ahn2014}. \cite{antolin2015} found coronal rain to be a multi-thermal phenomenon, likely with spatially thin transition between chromospheric and transition region (TR) temperatures.  Meanwhile, their observations suggests a relatively wider, yet still narrow, transition ($\sim0.5''$) from TR to coronal temperatures.

As neutral material does not experience a Lorentz force, how well do we expect our neutral helium observations to trace individual lines of magnetic force in the corona?  Similar to the H$\alpha$ line discussed by \cite{antolin2012sharp}, radiatively, the population of the \heI triplet lower level can be maintained via a recombination and rapid de-excitation cascade process \citep{avrett1994}, wherein free electrons are captured by available He \textsc{III} and He \textsc{II} ions.  Thus, the \heI 10830 \mbox{\AA} line formation is linked to the availability of ions tied to the magnetic field.  As \heI atoms are rapidly ionized within the megakelvin corona, this process suggests the neutral helium seen in the corona traces the cool components of the magnetic field.  Furthermore, the observation of \cite{ahn2014}, which are phenomenologically very similar to ours, showed neutral and ionized species to fall in unison. This, however, does not rule out possible partial ionization effects in rain conditions that decouple ionic and neutral species and permit cross field diffusion \citep[see, \textit{e.g.}, observations by][]{khomenko2016}.  For solar prominence material of representative parameters, \cite{gilbert2002} calculated the neutral helium cross-field downflow speed to be $\sim8 \times 10^{3}$ cm s$^{-1}$, while values reached $\sim 10$ km sec$^{-1}$ for prominence material of very low density ($\sim 10^{9}$ cm$^{-3}$), both of which are small fractions of the velocities observed here.

There is some evidence given by fully ionized simulations that the formation of condensates in the corona may induce variations in the magnetic field itself, and thus, our measurements of catastrophically cooled loops may not fully reflect the magnetic field configuration and intensity of its thermally stable predecessor.  With 2.5D thermodynamic magnetohydrodynamic simulations, \cite{fang2013} determined the pressure gradient brought on by the formation of small, but dense, coronal condensates introduces a force on par with the Lorentz force, and thus, the presence of the cool material can induce field variations locally, and within its neighborhood.  Due to the great EUV extinction along the loop we observe, the number densities along the cool loop are relatively high, and thus we expect there to be some small change is the magnetic flux density in the cool looped in comparison to its hot predecessor.  However, due to the correspondence between our measurements and previous measurements of hot coronal loops, we expect the effect to be small and not within the sensitivity of our measurements. 


\section{Summary}\label{sec:summary}

We have used neutral He I spectropolarimetry to study the vector magnetic field structure of a catastrophically cooled coronal loop up to heights of 70 Mm above the solar surface.  In congress with EUV stereoscopy, we have confirmed a number of expectations regarding the height stratification of coronal loops, namely that they are generally aligned with the magnetic field and exhibit a precipitous decrease in magnetic flux density with increasing height.  We believe the measurements demonstrated here to be a powerful means to diagnose the properties of coronal fine structure. 
 
In the near future, the Daniel K. Inouye Solar Telescope \citep{rimmele2015} will revolutionize our solar spectropolarimetric capabilities, in large part due to its substantial increase of light gathering capability (28 x that of the Dunn Solar Telescope).  The resulting higher signal-to-noise opportunities for the spatial and temporal scales discussed here, will greatly boost the sensitivity of this method to permit greater scrutiny of cool coronal loop magnetism, as well as open new diagnostic pathways, e.g., the use of the \heI polarization as a height diagnostic (see \cite{asensio_ramos2008} for discussion), and the use of the forbidden infrared hot coronal lines off-limb.


\acknowledgments

This work is dedicated to the memory of our friend and colleague, Mike Bradford.  SDO data are the courtesy of NASA/SDO and AIA and HMI science teams.  EUVI data are the courtesy of the STEREO Sun Earth Connection Coronal and Heliospheric Investigation (SECCHI) team.  FIRS has been developed by the Institute for Astronomy at the University of Hawai`i, jointly with the National Solar Observatory (NSO).  The FIRS project was funded by the National Science Foundation Major Research Instrument program, grant number ATM-0421582.  T.A.S. gratefully acknowledges colleagues at the High Altitude Obseratory for interesting discussions regarding this work during a scientific visit.  This work also benefited from discussions at the International Space Science Institute (ISSI) meetings on "Sub-arcsec Observations and Interpretation of the Chromosphere" and "Implications for coronal heating and magnetic fields from coronal rain observations and modelling."  This research has made use of NASA's Astrophysics Data System. \edit1{The authors thank the referee for a careful review and fruitful comments.}


\vspace{1mm}
\facilities{Dunn(FIRS), SDO(AIA), SDO(HMI), STEREO(EUVI-A)}


\bibliographystyle{aasjournal}
\bibliography{schad_manuscript_v5}

\begin{thebibliography}{}
\expandafter\ifx\csname natexlab\endcsname\relax\def\natexlab#1{#1}\fi

\bibitem[{{Ahn} {et~al.}(2014){Ahn}, {Chae}, {Cho}, {Song}, {Yang}, {Goode},
  {Cao}, {Park}, {Nah}, {Jang}, \& {Park}}]{ahn2014}
{Ahn}, K., {Chae}, J., {Cho}, K.-S., {et~al.} 2014, \solphys, 289, 4117

\bibitem[{{Allen} {et~al.}(1998){Allen}, {Bagenal}, \&
  {Hundhausen}}]{allen1998}
{Allen}, U.~A., {Bagenal}, F., \& {Hundhausen}, A.~J. 1998, in Astronomical
  Society of the Pacific Conference Series, Vol. 150, IAU Colloq. 167: New
  Perspectives on Solar Prominences, ed. D.~F. {Webb}, B.~{Schmieder}, \& D.~M.
  {Rust}, 290

\bibitem[{{Antolin} \& {Rouppe van der Voort}(2012)}]{antolin2012sharp}
{Antolin}, P., \& {Rouppe van der Voort}, L. 2012, \apj, 745, 152

\bibitem[{{Antolin} {et~al.}(2010){Antolin}, {Shibata}, \&
  {Vissers}}]{antolin2010}
{Antolin}, P., {Shibata}, K., \& {Vissers}, G. 2010, \apj, 716, 154

\bibitem[{{Antolin} {et~al.}(2015){Antolin}, {Vissers}, {Pereira}, {Rouppe van
  der Voort}, \& {Scullion}}]{antolin2015}
{Antolin}, P., {Vissers}, G., {Pereira}, T.~M.~D., {Rouppe van der Voort}, L.,
  \& {Scullion}, E. 2015, \apj, 806, 81

\bibitem[{{Antolin} {et~al.}(2012){Antolin}, {Vissers}, \& {Rouppe van der
  Voort}}]{antolin2012ondisk}
{Antolin}, P., {Vissers}, G., \& {Rouppe van der Voort}, L. 2012, \solphys,
  280, 457

\bibitem[{{Aschwanden}(2005)}]{aschwanden2005corona}
{Aschwanden}, M.~J. 2005, {Physics of the Solar Corona. An Introduction with
  Problems and Solutions (2nd edition)}

\bibitem[{{Aschwanden} {et~al.}(2012){Aschwanden}, {W{\"u}lser}, {Nitta}, \&
  {Lemen}}]{aschwanden2012}
{Aschwanden}, M.~J., {W{\"u}lser}, J.-P., {Nitta}, N., \& {Lemen}, J. 2012,
  \solphys, 281, 101

\bibitem[{{Aschwanden} {et~al.}(2008){Aschwanden}, {W{\"u}lser}, {Nitta}, \&
  {Lemen}}]{aschwanden2008}
{Aschwanden}, M.~J., {W{\"u}lser}, J.-P., {Nitta}, N.~V., \& {Lemen}, J.~R.
  2008, \apj, 679, 827

\bibitem[{{Asensio Ramos} {et~al.}(2008){Asensio Ramos}, {Trujillo Bueno}, \&
  {Landi Degl'Innocenti}}]{asensio_ramos2008}
{Asensio Ramos}, A., {Trujillo Bueno}, J., \& {Landi Degl'Innocenti}, E. 2008,
  \apj, 683, 542

\bibitem[{{Avrett} {et~al.}(1994){Avrett}, {Fontenla}, \&
  {Loeser}}]{avrett1994}
{Avrett}, E.~H., {Fontenla}, J.~M., \& {Loeser}, R. 1994, in IAU Symposium,
  Vol. 154, Infrared Solar Physics, ed. D.~M. {Rabin}, J.~T. {Jefferies}, \&
  C.~{Lindsey}, 35

\bibitem[{{Aznar Cuadrado} {et~al.}(2007){Aznar Cuadrado}, {Solanki}, \&
  {Lagg}}]{aznar_cuadrado2007}
{Aznar Cuadrado}, R., {Solanki}, S.~K., \& {Lagg}, A. 2007, in Modern solar
  facilities - advanced solar science, ed. F.~{Kneer}, K.~G. {Puschmann}, \&
  A.~D. {Wittmann}, 173

\bibitem[{{Breckinridge} \& {Hall}(1973)}]{breckinridge1973}
{Breckinridge}, J.~B., \& {Hall}, D.~N.~B. 1973, \solphys, 28, 15

\bibitem[{{Brosius} \& {White}(2006)}]{brosius2006}
{Brosius}, J.~W., \& {White}, S.~M. 2006, \apjl, 641, L69

\bibitem[{{Cargill} \& {Priest}(1980)}]{cargill1980}
{Cargill}, P.~J., \& {Priest}, E.~R. 1980, \solphys, 65, 251

\bibitem[{{Casini} {et~al.}(2012){Casini}, {Judge}, \&
  {Schad}}]{casini2012_fringes}
{Casini}, R., {Judge}, P.~G., \& {Schad}, T.~A. 2012, \apj, 756, 194

\bibitem[{{Casini} {et~al.}(2009){Casini}, {L{\'o}pez Ariste}, {Paletou}, \&
  {L{\'e}ger}}]{casini2009}
{Casini}, R., {L{\'o}pez Ariste}, A., {Paletou}, F., \& {L{\'e}ger}, L. 2009,
  \apj, 703, 114

\bibitem[{{Casini} {et~al.}(2003){Casini}, {L{\'o}pez Ariste}, {Tomczyk}, \&
  {Lites}}]{casini2003}
{Casini}, R., {L{\'o}pez Ariste}, A., {Tomczyk}, S., \& {Lites}, B.~W. 2003,
  \apjl, 598, L67

\bibitem[{{Centeno} {et~al.}(2010){Centeno}, {Trujillo Bueno}, \& {Asensio
  Ramos}}]{centeno2010}
{Centeno}, R., {Trujillo Bueno}, J., \& {Asensio Ramos}, A. 2010, \apj, 708,
  1579

\bibitem[{{Charbonneau}(1995)}]{charbonneau1995}
{Charbonneau}, P. 1995, \apjs, 101, 309

\bibitem[{{Chitta} {et~al.}(2016){Chitta}, {Peter}, \& {Young}}]{chitta2016}
{Chitta}, L.~P., {Peter}, H., \& {Young}, P.~R. 2016, \aap, 587, A20

\bibitem[{{Dove} {et~al.}(2011){Dove}, {Gibson}, {Rachmeler}, {Tomczyk}, \&
  {Judge}}]{dove2011}
{Dove}, J.~B., {Gibson}, S.~E., {Rachmeler}, L.~A., {Tomczyk}, S., \& {Judge},
  P. 2011, \apjl, 731, L1

\bibitem[{{Fang} {et~al.}(2013){Fang}, {Xia}, \& {Keppens}}]{fang2013}
{Fang}, X., {Xia}, C., \& {Keppens}, R. 2013, \apjl, 771, L29

\bibitem[{{Field}(1965)}]{field1965}
{Field}, G.~B. 1965, \apj, 142, 531

\bibitem[{{Gilbert} {et~al.}(2011){Gilbert}, {Kilper}, {Alexander}, \&
  {Kucera}}]{gilbert2011}
{Gilbert}, H., {Kilper}, G., {Alexander}, D., \& {Kucera}, T. 2011, \apj, 727,
  25

\bibitem[{{Gilbert} {et~al.}(2002){Gilbert}, {Hansteen}, \&
  {Holzer}}]{gilbert2002}
{Gilbert}, H.~R., {Hansteen}, V.~H., \& {Holzer}, T.~E. 2002, \apj, 577, 464

\bibitem[{{Gonz{\'a}lez Manrique} {et~al.}(2016){Gonz{\'a}lez Manrique},
  {Kuckein}, {Pastor Yabar}, {Collados}, {Denker}, {Fischer}, {G{\"o}m{\"o}ry},
  {Diercke}, {Bello Gonz{\'a}lez}, {Schlichenmaier}, {Balthasar}, {Berkefeld},
  {Feller}, {Hoch}, {Hofmann}, {Kneer}, {Lagg}, {Nicklas}, {Orozco Su{\'a}rez},
  {Schmidt}, {Schmidt}, {Sigwarth}, {Sobotka}, {Solanki}, {Soltau}, {Staude},
  {Strassmeier}, {Verma}, {Volkmer}, {von der L{\"u}he}, \&
  {Waldmann}}]{gonzalez_manrique2016}
{Gonz{\'a}lez Manrique}, S.~J., {Kuckein}, C., {Pastor Yabar}, A., {et~al.}
  2016, ArXiv e-prints, arXiv:1603.00679

\bibitem[{{Howard} {et~al.}(1990){Howard}, {Harvey}, \& {Forgach}}]{howard1990}
{Howard}, R.~F., {Harvey}, J.~W., \& {Forgach}, S. 1990, \solphys, 130, 295

\bibitem[{{Jaeggli} {et~al.}(2010){Jaeggli}, {Lin}, {Mickey}, {Kuhn}, {Hegwer},
  {Rimmele}, \& {Penn}}]{jaeggli2010}
{Jaeggli}, S.~A., {Lin}, H., {Mickey}, D.~L., {et~al.} 2010, \memsai, 81, 763

\bibitem[{{Judge} {et~al.}(2001){Judge}, {Casini}, {Tomczyk}, {Edwards}, \&
  {Francis}}]{judge2001}
{Judge}, P.~G., {Casini}, R., {Tomczyk}, S., {Edwards}, D.~P., \& {Francis}, E.
  2001, NCAR/TN-446+STR, doi:10.5065/D6222RQH

\bibitem[{{Kaiser}(2005)}]{kaiser2005}
{Kaiser}, M.~L. 2005, Advances in Space Research, 36, 1483

\bibitem[{{Kamio} {et~al.}(2011){Kamio}, {Peter}, {Curdt}, \&
  {Solanki}}]{kamio2011}
{Kamio}, S., {Peter}, H., {Curdt}, W., \& {Solanki}, S.~K. 2011, \aap, 532, A96

\bibitem[{Khomenko {et~al.}(2016)Khomenko, Collados, \& Díaz}]{khomenko2016}
Khomenko, E., Collados, M., \& Díaz, A.~J. 2016, The Astrophysical Journal,
  823, 132

\bibitem[{{Kjeldseth-Moe} \& {Brekke}(1998)}]{kjeldseth_moe_1998}
{Kjeldseth-Moe}, O., \& {Brekke}, P. 1998, \solphys, 182, 73

\bibitem[{{Kleint} {et~al.}(2014){Kleint}, {Antolin}, {Tian}, {Judge}, {Testa},
  {De Pontieu}, {Mart{\'{\i}}nez-Sykora}, {Reeves}, {Wuelser}, {McKillop},
  {Saar}, {Carlsson}, {Boerner}, {Hurlburt}, {Lemen}, {Tarbell}, {Title},
  {Golub}, {Hansteen}, {Jaeggli}, \& {Kankelborg}}]{kleint2014}
{Kleint}, L., {Antolin}, P., {Tian}, H., {et~al.} 2014, \apjl, 789, L42

\bibitem[{{Kuckein} {et~al.}(2012{\natexlab{a}}){Kuckein}, {Mart{\'{\i}}nez
  Pillet}, \& {Centeno}}]{kuckein2012mag}
{Kuckein}, C., {Mart{\'{\i}}nez Pillet}, V., \& {Centeno}, R.
  2012{\natexlab{a}}, \aap, 539, A131

\bibitem[{{Kuckein} {et~al.}(2012{\natexlab{b}}){Kuckein}, {Mart{\'{\i}}nez
  Pillet}, \& {Centeno}}]{kuckein2012}
---. 2012{\natexlab{b}}, \aap, 542, A112

\bibitem[{{Kurucz}(1995)}]{kurucz1995}
{Kurucz}, R.~L. 1995, in Astronomical Society of the Pacific Conference Series,
  Vol.~78, Astrophysical Applications of Powerful New Databases, ed. S.~J.
  {Adelman} \& W.~L. {Wiese}, 205

\bibitem[{{Lagg}(2007)}]{lagg2007a}
{Lagg}, A. 2007, Advances in Space Research, 39, 1734

\bibitem[{{Lagg} {et~al.}(2004){Lagg}, {Woch}, {Krupp}, \&
  {Solanki}}]{lagg2004}
{Lagg}, A., {Woch}, J., {Krupp}, N., \& {Solanki}, S.~K. 2004, \aap, 414, 1109

\bibitem[{{Lagg} {et~al.}(2007){Lagg}, {Woch}, {Solanki}, \&
  {Krupp}}]{lagg2007b}
{Lagg}, A., {Woch}, J., {Solanki}, S.~K., \& {Krupp}, N. 2007, \aap, 462, 1147

\bibitem[{{Landi Degl'Innocenti} \& {Landolfi}(2004)}]{landi_2004}
{Landi Degl'Innocenti}, E., \& {Landolfi}, M., eds. 2004, Astrophysics and
  Space Science Library, Vol. 307, {Polarization in Spectral Lines}

\bibitem[{Lee \& Schachter(1980)}]{lee1980}
Lee, D.~T., \& Schachter, B.~J. 1980, International Journal of Computer {\&}
  Information Sciences, 9, 219

\bibitem[{{Lemen} {et~al.}(2011){Lemen}, {Title}, {Akin}, {Boerner}, {Chou},
  {Drake}, {Duncan}, {Edwards}, {Friedlaender}, {Heyman}, {Hurlburt}, {Katz},
  {Kushner}, {Levay}, {Lindgren}, {Mathur}, {McFeaters}, {Mitchell}, {Rehse},
  {Schrijver}, {Springer}, {Stern}, {Tarbell}, {Wuelser}, {Wolfson}, {Yanari},
  {Bookbinder}, {Cheimets}, {Caldwell}, {Deluca}, {Gates}, {Golub}, {Park},
  {Podgorski}, {Bush}, {Scherrer}, {Gummin}, {Smith}, {Auker}, {Jerram},
  {Pool}, {Soufli}, {Windt}, {Beardsley}, {Clapp}, {Lang}, \&
  {Waltham}}]{lemen2011}
{Lemen}, J.~R., {Title}, A.~M., {Akin}, D.~J., {et~al.} 2011, \solphys, 172

\bibitem[{{Lin} {et~al.}(2004){Lin}, {Kuhn}, \& {Coulter}}]{lin2004}
{Lin}, H., {Kuhn}, J.~R., \& {Coulter}, R. 2004, \apjl, 613, L177

\bibitem[{{Lin} {et~al.}(2006){Lin}, {Martin}, \& {Engvold}}]{lin_martin2006}
{Lin}, Y., {Martin}, S.~F., \& {Engvold}, O. 2006, in Bulletin of the American
  Astronomical Society, Vol.~38, AAS/Solar Physics Division Meeting 37, 219

\bibitem[{{Liu} \& {Lin}(2008)}]{liu_lin2008}
{Liu}, Y., \& {Lin}, H. 2008, \apj, 680, 1496

\bibitem[{{Moschou} {et~al.}(2015){Moschou}, {Keppens}, {Xia}, \&
  {Fang}}]{moschou2015}
{Moschou}, S.~P., {Keppens}, R., {Xia}, C., \& {Fang}, X. 2015, Advances in
  Space Research, 56, 2738

\bibitem[{{Oliver} {et~al.}(2016){Oliver}, {Soler}, {Terradas}, \&
  {Zaqarashvili}}]{oliver2016}
{Oliver}, R., {Soler}, R., {Terradas}, J., \& {Zaqarashvili}, T.~V. 2016, \apj,
  818, 128

\bibitem[{{Orange} {et~al.}(2014){Orange}, {Oluseyi}, {Chesny}, {Patel},
  {Champey}, {Hesterly}, {Anthony}, \& {Treen}}]{orange2014}
{Orange}, N.~B., {Oluseyi}, H.~M., {Chesny}, D.~L., {et~al.} 2014, \solphys,
  289, 1901

\bibitem[{{Penn}(2000)}]{penn2000}
{Penn}, M.~J. 2000, \solphys, 197, 313

\bibitem[{{Pesnell} {et~al.}(2012){Pesnell}, {Thompson}, \&
  {Chamberlin}}]{pesnell2012}
{Pesnell}, W.~D., {Thompson}, B.~J., \& {Chamberlin}, P.~C. 2012, \solphys,
  275, 3

\bibitem[{{Rimmele} {et~al.}(2015){Rimmele}, {McMullin}, {Warner}, {Craig},
  {Woeger}, {Tritschler}, {Cassini}, {Kuhn}, {Lin}, {Schmidt}, {Berukoff},
  {Reardon}, {Goode}, {Knoelker}, {Rosner}, {Mathioudakis}, \& {DKIST
  TEAM}}]{rimmele2015}
{Rimmele}, T., {McMullin}, J., {Warner}, M., {et~al.} 2015, IAU General
  Assembly, 22, 2255176

\bibitem[{{Rimmele} {et~al.}(2004){Rimmele}, {Richards}, {Hegwer}, {Fletcher},
  {Gregory}, {Moretto}, {Didkovsky}, {Denker}, {Dolgushin}, {Goode},
  {Langlois}, {Marino}, \& {Marquette}}]{rimmele2004}
{Rimmele}, T.~R., {Richards}, K., {Hegwer}, S., {et~al.} 2004, in Society of
  Photo-Optical Instrumentation Engineers (SPIE) Conference Series, Vol. 5171,
  Society of Photo-Optical Instrumentation Engineers (SPIE) Conference Series,
  ed. S.~{Fineschi} \& M.~A. {Gummin}, 179--186

\bibitem[{{Sasso} {et~al.}(2011){Sasso}, {Lagg}, \& {Solanki}}]{sasso2011}
{Sasso}, C., {Lagg}, A., \& {Solanki}, S.~K. 2011, \aap, 526, A42

\bibitem[{{Schad}(2014)}]{schad2014}
{Schad}, T.~A. 2014, \solphys, 289, 1477

\bibitem[{{Schad} {et~al.}(2013){Schad}, {Penn}, \& {Lin}}]{schad2013}
{Schad}, T.~A., {Penn}, M.~J., \& {Lin}, H. 2013, \apj, 768, 111

\bibitem[{{Schad} {et~al.}(2015){Schad}, {Penn}, {Lin}, \&
  {Tritschler}}]{schad2015}
{Schad}, T.~A., {Penn}, M.~J., {Lin}, H., \& {Tritschler}, A. 2015, \solphys,
  290, 1607

\bibitem[{{Schmidt} {et~al.}(2000){Schmidt}, {Muglach}, \&
  {Kn{\"o}lker}}]{schmidt2000}
{Schmidt}, W., {Muglach}, K., \& {Kn{\"o}lker}, M. 2000, \apj, 544, 567

\bibitem[{{Schrijver}(2001)}]{schrijver2001}
{Schrijver}, C.~J. 2001, \solphys, 198, 325

\bibitem[{{Socas-Navarro} {et~al.}(2005){Socas-Navarro}, {Trujillo Bueno}, \&
  {Landi Degl'Innocenti}}]{socas_navarro2005}
{Socas-Navarro}, H., {Trujillo Bueno}, J., \& {Landi Degl'Innocenti}, E. 2005,
  \apjs, 160, 312

\bibitem[{{Tomczyk} {et~al.}(2007){Tomczyk}, {McIntosh}, {Keil}, {Judge},
  {Schad}, {Seeley}, \& {Edmondson}}]{tomczyk2007}
{Tomczyk}, S., {McIntosh}, S.~W., {Keil}, S.~L., {et~al.} 2007, Science, 317,
  1192

\bibitem[{{Tomczyk} {et~al.}(2008){Tomczyk}, {Card}, {Darnell}, {Elmore},
  {Lull}, {Nelson}, {Streander}, {Burkepile}, {Casini}, \&
  {Judge}}]{tomcyzk2008}
{Tomczyk}, S., {Card}, G.~L., {Darnell}, T., {et~al.} 2008, \solphys, 247, 411

\bibitem[{{Trujillo Bueno} {et~al.}(2002){Trujillo Bueno}, {Landi
  Degl'Innocenti}, {Collados}, {Merenda}, \& {Manso
  Sainz}}]{trujillo_bueno_2002}
{Trujillo Bueno}, J., {Landi Degl'Innocenti}, E., {Collados}, M., {Merenda},
  L., \& {Manso Sainz}, R. 2002, \nat, 415, 403

\bibitem[{{Vissers} \& {Rouppe van der Voort}(2012)}]{vissers2012}
{Vissers}, G., \& {Rouppe van der Voort}, L. 2012, \apj, 750, 22

\end{thebibliography}


\end{document}